\documentclass{raa} 
\usepackage{graphicx}
\usepackage{epsf}
\usepackage{color}
\usepackage{supertabular}
\usepackage{hyperref}
\usepackage{amsmath}
\usepackage{longtable}
\usepackage{tabularx}
\usepackage{booktabs}
\usepackage{array}
\usepackage{amssymb}
\usepackage{threeparttable}
\usepackage{booktabs}
\usepackage{times}

\newcommand{\logg}{log\,$g$}

\begin{document}

   \title{On the metallicity gradients of the Galactic disk as revealed by LSS-GAC red clump stars}

   \volnopage{Vol.0 (200x) No.0, 000--000}      %%preserved for Editor. DOn't remove!
   \setcounter{page}{1}          %%starting page, preserved for Editor. DOn't remove!
   \author{Yang Huang
      \inst{1}
   \and Xiao-Wei Liu
      \inst{1,2}
   \and Hua-Wei Zhang
      \inst{1,2}
     \and Hai-Bo Yuan
     \inst{2}
      \and Mao-Sheng Xiang
      \inst{1}  
     \and Bing-Qiu Chen
      \inst{2}  
      \and Juan-Juan Ren
      \inst{1}
      \and Ning-Chen Sun
      \inst{1}
       \and Chun Wang
      \inst{1}
      \and{Yong Zhang}
      \inst{3}
      \and{Yong-Hui Hou}
      \inst{3}
      \and{Yue-Fei Wang}
      \inst{3}
      \and{Ming Yang}
      \inst{4}
   }

   \institute{Department of Astronomy, Peking University, Beijing 100871, China; {\it yanghuang@pku.edu.cn {\rm (YH)};  x.liu@pku.edu.cn {\rm (XWL)}; zhanghw@pku.edu.cn {\rm(HWZ)}}\\   
        \and
             Kavli Institute for Astronomy and Astrophysics, Peking University, Beijing 100871, China\\
          \and
          Nanjing Institute of Astronomical Optics \& Technology, National Astronomical Observatories, Chinese Academy of Sciences, Nanjing 210042, China\\
          \and 
          Key Laboratory of Optical Astronomy, National Astronomical Observatories, Chinese Academy of Sciences, Beijing 100012, China \\
}

\date{Received~~2009 month day; accepted~~2009~~month day}

\abstract{ Using a sample of over 70,\,000 red clump (RC)  stars with $5$\,--\,$10$\,\% distance accuracy selected from the LAMOST Spectroscopic Survey of the Galactic Anti-center (LSS-GAC), we study the radial and vertical gradients of the Galactic disk(s) mainly in the anti-center direction, covering a significant disk volume of projected Galactocentric radius $7\,\leq\,R_{\rm GC}\,\leq\,14$\,kpc and height from the Galactic midplane $0\,\leq\,|Z|\,\leq\,3$\,kpc.
Our analysis  shows that both the radial and vertical metallicity gradients are negative across much of the disk volume probed, and exhibit significant spatial variations.
Near the solar circle ($7 \leq R_{\rm GC} \leq 11.5$\,kpc), the radial gradient has a moderately steep, negative slope of $-0.08$\,dex\,kpc$^{-1}$ near the midplane ($|Z| < 0.1$\,kpc), and the slope  flattens with increasing  $|Z|$.
In the outer disk ($11.5\,<\,R_{\rm GC}\,\leq\,14$\,kpc), the radial gradients have an essentially  constant, much less steep slope of $-0.01\,$\,dex\,kpc\,$^{-1}$ at all heights above the plane, suggesting that the outer disk may have experienced an evolution path different from that of the inner disk.
The vertical  gradients are  found to  flatten largely with increasing $R_{\rm GC}$. 
However,  the vertical gradient of the  lower disk ($0\,\leq\,|Z|\,\leq\,1$\,kpc) is found to flatten with $R_{\rm GC}$ quicker than that of the upper disk ($1\,<\,|Z|\,\leq\,3$\,kpc).
Our results should provide strong constraints on the theory of disk formation and evolution, as well as the underlying physical processes that shape the disk (e.g. gas flows, radial migration, internal and external perturbations).
}
\authorrunning {Y. Huang, X.-W. Liu, H.-W. Zhang et al.}

%\begin{keywords}
%Galaxy: abundance -- Galaxy: disk -- Galaxy: evolution -- Galaxy: formation -- techniques: spectroscopic
%\end{keywords}

\titlerunning{Galactic metallicity gradients}  % title_head in odd pages

   \maketitle

\section{Introduction} 
%Overall introduction
The distribution of metals in the disk of a galaxy is the consequence of a variety of complicated processes, including star formation and evolution, gas flows and stellar migration, non-axisymmetric perturbations (e.g. by the central bar, spiral arms), and interactions and accretions.
The stellar metallicity distribution in a disk galaxy including our own Milky Way typically exhibits a (negative) gradient both in the radial and vertical direction.
Measurements of (radial and vertical) metallicity gradients are therefore of great importance to constrain the formation and evolution of galaxies, especially of the Milky Way.

%Firstly: observation
%a) various tracers & general introduction:
The radial metallicity gradient of the Milky Way disk(s) has been measured using various tracers, including classical Cepheids (e.g. Andrievsky et al. 2002a,b,c, 2004; Luck et al. 2003, 2006, 2011; Kovtyukh et al. 2005; Yong et al. 2006; Lemasle et al. 2007, 2008; Pedicelli et al. 2009, 2010; Luck\,\&\,Lambert 2011;  Genovali et al. 2013, 2014; Martin et al. 2015), H~{\sc ii} regions (e.g. Deharveng et al. 2000; Quireza et al. 2006; Rudolph et al. 2006; Balser et al. 2011), O/B-type stars (e.g. Smartt \& Rolleston 1997; Gummersbach et al. 1998; Rolleston 2000; Daflon \& Cunha 2004; Daflon et al. 2009),  open clusters (OCs; Friel et al. 2002, 2010; Chen et al. 2003; Yong et al. 2005, 2012; Carraro et al. 2007; Jacobson et al. 2008, 2009, 2011a,b; Magrini et al. 2009, 2010; Carrera \& Pancino 2011; Frinchaboy et al. 2013) and planetary nebulae (PNe; Henry et al. 2004, 2010; Maciel, Lago\,\&\,Costa 2005; Stanghellini et al. 2006; Perinotto \& Morbidelli 2006; Stanghellini\,\&\,Haywood 2010).
The above tracers are mainly close to the midplane  and the sample sizes are generally limited to a few hundred targets.
Recently, with the advent of large scale spectroscopic surveys of the Milky Way,  radial gradients at different heights ($Z$'s) from the midplane and vertical gradients at different projected Galactocentric radii ($R_{\rm GC}$'s) have been measured with large numbers ($\sim$\,tens of thousands) of field main-sequence/giant stars, such as main-sequence turnoff stars used by Cheng et al. (2012) from the SDSS/SEGUE (York et al. 2000; Yanny et al. 2009), both dwarf and giant stars used by Boeche et al. (2013, 2014) from the RAVE survey (Steinmetz et al. 2006), red giants used by Hayden et al. (2014) from the SDSS-III/APOGEE (Allende Prieto et al. 2008; Ahn et al. 2014).
{ Those tracers retain the chemical composition of the interstellar medium at the time of their birth.
Hence,  tracers of different ages  probe metallicity gradients of the disk at different epoches.}
Near the solar circle ($7<$\,$R_{\rm GC}$\,$< 10$\,kpc), the measured radial gradients of the disk measured with different tracers range from $-0.01$ to $-0.09$\,dex\,kpc$^{-1}$ (e.g. Cheng et al. 2012; Lemasle et al. 2013) and the vertical gradients range from $-0.07$ to $-0.30$\,dex\,kpc$^{-1}$ (e.g. Katz et al. 2011; Chen et al. 2011; Kordopatis et al. 2011; Schlesinger et al. 2012, 2014).
The negative radial gradients are generally explained with the so-called {\it inside-out} chemical evolution model (e.g, Larson et al. 1976; Matteucci\,\&\,Francois 1989; Chiappini et al. 1997, 2001; Hou et al. 2000; Prantzos\,\&\,Boissier 2000; Moll\'a\,\&\,D\'iaz 2005; Fu et al. 2009) in which the inner region of the disk is chemically enriched faster than the outer region.

%b) the behaviours
However, there has been evidence that even close to the disk midplane, a simple linear relation may not be sufficient to  describe the full variations  of metallicity over the full range of  $R_{\rm GC}$.
Using classical Cepheids, several studies (e.g. Andrievsky et al. 2002b; Pedicelli et al. 2009, 2010; Luck \& Lambert  2011; Genovali et al. 2013) have discovered that the radial gradient in the inner disk ($4$\,$\leq$\,$R_{\rm GC}$\,$\leq$\,$7$\,kpc) is steeper than that near the solar circle. 
On the other hand, a flat, nearly zero radial gradient in the inner disk ($R_{\rm GC} \leq 6$\,kpc) has been found recently by Hayden et al. (2014)  using APOGEE red giants. 
In the outer disk ($R_{\rm GC}$\,$>$\,$10$\,--\,$12$\,kpc), most studies based on OCs find that the radial gradient tends to be close to zero, and is flatter than that near the solar circle.   
Similar flat gradients in the outskirts of the disk are also reported by studies using other tracers, including H~{\sc ii} regions (Vilchez \& Esteban 1996), PNe (Costa et al. 2004) and classical Cepheids (Andrievsky et al. 2002c, 2004; Luck et al. 2003; Lemasle et al. 2008).
However, it is still not clear that those observed flat gradients are real or simply due to the coarse sampling since almost all the aforementioned  studies are based on samples containing only few dozens of targets in the outer disk.
More recently, more classical Cepheids have been found in the outer disk. 
The newly radial gradient seems to have a value comparable to that near the solar circle, lending further evidence that the previously reported  flat gradients in the outer disk may be indeed artifact caused by the effects of poor sampling (e.g. Luck et al. 2011; Luck\,\&\,Lambert 2011). 
From the theoretical point of view, a flat radial gradient in the outer disk can be explained by  the natural outcome of disk formation and evolution (e.g. Carraro et al. 2007), { the mixing effects due to the presence of non-axisymmetric structures (e.g. central bar, long-lived spiral arm;  e.g. Andrievsky et al. 2004; Scarano\,\&\,L\'epine 2013), or by a merge event (e.g. Yong 2005).} 

%structure evolution
Out of the midplane, the radial gradients are found to flatten or even reverse to a positive slope as with the height from the midplane increases (Allende Prieto et al. 2006; Juri\'c et al. 2008; Carrell et al. 2012; Cheng et al. 2012; Hayden et al. 2014).
The flattening of the radial gradient are possibly linked to the distributions of different stellar populations of the Galactic disk(s) and may possibly be explained by the different levels of mixture of thin- and thick-disk stars as height varies (e.g. Juri\'c et al. 2008). 
The vertical gradients have also been found to  show spatial variations in the sense that the gradients flatten with increasing $R_{\rm GC}$ (e.g. Hayden et al. 2014).

%time evolution
{The evolution  of radial gradient is of utter  importance for an understanding of the formation and evolution of the Galactic disk(s).
However, the results of existing studies of temporal variations of the radial gradient of the disk are far from being unequivocal, both observationally and theoretically.}
Most studies based on OCs have found a flattening radial gradient with cosmic time by dividing the OCs into different age bins.
The results from PNe are less clear:  the flattening, steepening and unchanged are all reported in the literature (e.g. Maciel, Lago\,\&\,Costa 2005; Stanghellini\,\&\,Haywood 2010; Henry et al. 2010).
It should be emphasized that all the tracers (OCs/PNe) used in the above analyses are younger than about 5\,--\,7\,Gyr (see Maciel, Lago \& Costa 2005).
The situation of possible  radial gradient evolution at the early epochs of disk formation is still unclear.
Theoretically, different chemical evolution models predict different evolution of the radial gradient: flattening with cosmic time (e.g. Moll\'a et al. 1997; Portinari \& Chiosi 1999; Hou et al. 2000) or steepening with cosmic time (e.g. Tosi 1988; Chiappini et al. 1997, 2001).   
More recently, the radial migration has been suggested to be an important process in regulating the evolution of  radial gradient and thus has to be included in the chemical evolution models (e.g. Sch\"onrich \& Binney 2009; Loebman et al. 2011; Kubryk et al. 2013; Minchev et al. 2013).
Generally, the radial gradients of the old populations will be strongly affected and flattened by the migration.

%The current challenges
Compared to measurements of the exact slope of the metallicity gradient,
studies of the spatial and temporal variations of the gradients are more fundamental for understanding the formation and evolution of the Galactic disk(s) since the variations contain information about the underlying physical processes that regulate the disk formation and evolution at different times, and in different regions of the disk.
However, as described above, the currently available measurements of the spatial and temporal variations of the gradients are still quite controversial and in doubt.
{
This is possibly because most of the previous studies have not considered the temporal variations of gradients when deriving the spatial variations of gradients, or, the spatial variations of gradients when deducing the temporal variations of gradients ,  given the following considerations.
%considering that the spatial and temporal variations are coupled each other
%Firstly, the spatial distributions of different tracers (ages) change a lot that the younger population are closer to the midplane while the older one are much more spread out.
Firstly, the metallicity gradients in different regions of the disk may vary  significantly, due to, for example, the effects of  the non-axisymmetric perturbations or  merge events.
Secondly, as discussed earlier, different tracers probe metallicity gradients at different evolutionary epochs of the Galaxy. 
Finally, the potential biases caused by target selection effects are not always fully explored in many  of the existing studies. 
This last  effect is, at least partly, responsible for the differing  results yielded by different studies.
%The various results about the spatial and temporal variations will be yielded 
%Most of the previous studies have not considered the temporal variations in deriving spatial variations and vice versa; secondly, the potential target selection biases are not explored for most of the studies. 
}
%Actually, the spatial and temporal variations are coupled each other given the fact that }
%This is possible because the previous studies have not considered the temporal variations in deriving spatial variations and vice versa due to the limited size and spatial coverages, as well as the poorly constrained ages, for most sample used.
%Moreover,  the potential target selection biases are not explored for most of the studies. 
A large-scale spectroscopic survey that targets large numbers of stars over the full age range of the Galaxy and covers a sufficiently  large volume of the Galactic disk(s)  with sufficient sampling number density for all stellar populations concerned, is thus required to address the above problems properly and eventually draw a clear picture about how the Galactic disk(s) forms and evolves chemo-dynamically. 
 
%LSS-GAC
The Large Sky Area Multi-Object Fiber Spectroscopic Telescope\footnote{LAMOST is a 4 metre quasi-meridian reflecting Schmidt telescope equipped with 4000 fibers, each of an angular diameter of 3.3 arcsec projected on the sky, distributed in a circular field of view of 5$^{\circ}$ in diameter (Cui et al. 2012).} (LAMOST, also named the Guoshoujing Telescope) Spectroscopic Survey of the Galactic Anti-center (LSS-GAC; Liu et al. 2014; Yuan et al. 2015), as a major component of the on-going {LAMOST Experiment for Galactic Understanding and Exploration (LEGUE; Deng et al. 2012)}, aims to collect low resolution ($R$\,$\sim$\,$1800$), optical ($\lambda\lambda$\,3800\,--\,9000) spectra under dark and grey lunar conditions for a statistically complete sample of over 3 million stars of magnitudes $14.0$\,$\leq$\,$r$\, $\leq$\, $17.8$\,mag (down to $18.5$\,mag for limited fields), in a contiguous sky area of $\sim$\,3400\,sq.deg., centred on the Galactic anticentre (GAC), covering Galactic longitudes $150$\,$<$\,$l$\,$ <$\,$210^{\circ}$ and latitudes $|b|$\,$<$\,$30^{\circ}$. 
Over 1.5 million spectra of very bright stars ($9$\,$<$\,$r$\,$<$\,$14.0$\,mag) will also be obtained under bright lunar conditions. 
The survey will deliver spectral classifications, values of stellar radial velocity $V_{\rm r}$ and stellar atmospheric parameters (effective temperature $T_{\rm eff}$, surface gravity log\,$g$, metallicity [Fe/H], [$\alpha$/Fe] and [C/Fe]) from the collected spectra.
As designed, the survey will provide a large number of stars, of the order of a few millions, with a contiguous sky coverage of a large area of the Galactic disk(s) with a notably sampling density (hundreds of stars per squared degree) yet using a simple but nonetheless nontrivial target selection algorithms (random selection in the colour--magnitude diagram, i.e. $r$\,versus\,$g-r$ and $r$\,versus\,$r-i$). The survey will thus enable simultaneous measurements of the radial and vertical metallicity gradients across a large volume of the Galactic disk(s). 
The LSS-GAC regular survey was initiated  in September 2012.
It is expected to last for five years.

Based on the large stellar spectroscopic samples obtained by the LSS-GAC, we have started a series of investigations to systematically study the spatial and temporal variations of metallicity gradients of the Galactic disk(s).
Xiang et al. (this volume, hereafter Paper\,I) examine the spatial and temporal variations (in both radial and vertical directions) of metallicity gradients for $7.5<R_{\rm GC}<13.5$\,kpc and $|Z|<2.5$\,kpc using a sample of nearly 0.3 million main-sequence turnoff (MSTO) stars of ages ranging from 2 to over 11 Gyr, selected from the LSS-GAC DR2 (Xiang et al. 2015a, in preparation).  
%{ With the isochrone age determinations for each individual MSTO star,  significant temporal variations are found  by Paper I in both the radial and vertical gradients and suggest a two-phase formation of the Galactic disk.}
The current study, the second one of this series, selects over 70, 000 red clump (RC) giants { of intermediate- to old-ages} also from the LSS-GAC DR2.
The bright intrinsic luminosities of RC stars allow us to investigate  the (radial and vertical) gradients over a  slightly larger volume of the disk, covering  $7.0$\,$<$\,$R_{\rm GC}$\,$<$\,$14.0$\,kpc and $|Z|$\,$<$\,$3.0$\,kpc,  than possible with the MSTO star sample  used by Paper\,I. 
The current  sample is thus particularly suitable for the study of the outer disk and provides a unique opportunity to address the question whether the radial gradient of the outer disk is flatter than that near the solar circle as described above.
%To examine independently the temporal variations of the radial metallicity gradient, we have also collected two other independent  samples with accurate age determinations from the literature (see Section\,2.3 for details)  and included their analysis in the current study.
The paper is organized as follows. We describe in  detail the selection of  RC sample  in Section\,2.
We present our results in Section\,3, followed by discussions in Section\,4.
Conclusions are given in  Section\,5.

\section{Data and  Sample Selection}
\subsection{Data}
The data used in the current study are mainly collected during the Pilot (Sep.\,2011\,--\,Jun.\,2012 ) and first-two-year (Sept.\,2012\,--\,Jun.\,2014) Regular Surveys of LSS-GAC.
The stellar radial velocity $V_{\rm r}$ and stellar atmospheric parameters ($T_{\rm eff}$, log\,$g$, [Fe/H]) are determined with the LAMOST Stellar Parameter Pipeline at Peking University (LSP3; Xiang et al. 2015b,c), and have an overall accuracy of 5--10\,km\,s$^{-1}$, 150\,K, 0.25\,dex, 0.15\,dex, respectively, for a spectral signal-to-noise ratio per pixel ($\sim$\,1.07\,\AA) at 4650\,\AA, S/N (4650\,\AA )\,$\geq$\,10.
Stellar parameters, together with values of interstellar extinction and stellar distance estimated with a variety of techniques of over 0.7 million stars observed in the Pilot and first-year Regular Surveys are now publicly available as the first data release of value-added catalog of LSS-GAC (LSS-GAC DR1; Yuan et al. 2015).
The data of another 0.7 million stars with S/N (4650\,\AA )\,$\geq$\,10 from the second-year Regular Survey will soon be publicly available as the second release of value-added catalog of LSS-GAC (Xiang et al. 2015a, in preparation).
In addition to the LSS-GAC data, the current study has also included about 80,\,000 stars from { the spheroid parts ($|b|\,>\,20^{\circ}$) of the LEGUE} and 50,\,000 stars from LAMOST--$Kepler$ Field Survey (De Cat et al. 2014).
The latter are particularly useful for the RC selections (see Section\,2.2).
The additional spectra from the latter two surveys are also processed in the same way as for the LSS-GAC sources with the LSP3.

\subsection{The RC  sample}
%Briefly introduction of RC
\subsubsection{An overview of RC stars}
The helium burning RC stars\footnote{{ RC stars of interest  here refer to those in which the helium burning is  ignited following a helium flash that breaks the degeneracy of the helium core.}} are an easily recognizable feature in the colour-magnitude diagram (CMD) of intermediate- to old-age, low-mass but metal-rich stellar populations.
RC stars are considered as standard candles since their absolute luminosities are fairly independent of the stellar chemical composition and age (e.g. Cannon 1970; Paczy\'nski\,\&\,Stanek 1998).
With the accurate calibration of absolute magnitudes of several hundred local RC stars using the {\it Hipparcos} (ESA\,1997)  parallaxes becoming available, 
RC stars have become popular distance indicators and been widely used to obtain precise distances to the Galactic Center (Paczy\'nski\,\&\,Stanek 1998) and the Local Group of galaxies (e.g. LMC by Laney et al. 2012; M31 by Stanek \& Granvich 1998).  
The key of this method is a notable over-density feature easily identifiable in the CMD of stars in those distant objects due to the presence of a large number of RC stars at nearly the same distance. 

However, identifying individual RC stars amongst the numerous field stars is not an easy task.
Field RC stars spread over a wide range of distances and there is no apparent over-density in  the CMD of field stars that can be used to single out the RC members.
Fortunately, high-resolution spectroscopic analysis of nearby RC stars show that they distribute in a relative ``small box'' in the $T_{\rm eff}$\,--\,log\,$g$ diagram (i.e. HR diagram) of $4800$\,$\leq$\,$T_{\rm eff}$\,$\leq$\,$5200$\,K and $2.0$\,$\leq$\,${\rm log}$\,$g$\,$\leq$\,$3.0$ (e.g. Puzeras et al. 2010).
This is simply because  $T_{\rm eff}$ is an excellent proxy of stellar color whereas  log\,$g$ is sensitive to the stellar absolute luminosity.
Recently, with values of $T_{\rm eff}$ and log\,$g$ available from large-scale spectroscopic surveys  for large numbers of field stars, it has become feasible to select a large number of  field RC candidates.
RC candidates thus selected have been  widely used to study the metallicity gradients and stellar kinematics of the Galactic disk(s) (e.g. Siebert et al. 2011; Bilir et al. 2012; Williams et al. 2013; Bienaym\'e et al. 2014).
However, stars falling inside  that ``small box'' are not purely RC stars and have significant ($\sim 60$\%; Williams et al. 2013) contamination from red giant branch (RGB).
The differences in absolute magnitudes between RC and RGB stars can be larger than 1\,mag and this can lead to large systematic errors in distances of the selected  RC sample stars.
More recently, a new method has been proposed by Bovy et al. (2014; hereafter B14) to select a clean  RC sample from large-scale spectroscopic data.
The method first separates the RC-like and RGB stars using cuts on a metallicity-dependent $T_{\rm eff}$\,--\,log\,$g$ diagram, assisted by theoretical using stellar isochrones and calibrated using {\em Kepler} high quality asteroseismic log\,$g$ (e.g. Creevey et al. 2013) as well as high precision stellar atmospheric parameters from the APOGEE survey (Pinsonneault et al. 2014).
Secondly, secondary RC stars (i.e. high mass helium burning stars ignited nondegeneratedly) are then removed from the RC-like stars via cuts on a metallicity ($Z$)\,--\,color $(J-K_{\rm s})_{0}$ diagram.
The expected purity of the final RC sample is $\geq$\,$93$\,\% and the typical distance errors are within 5\,--\,10\,\%.
The key point of this new method in selecting a clean RC sample is accurate values of surface gravity log\,$g$.   
Unfortunately, it is difficult to apply this method to the current LSS-GAC data given the relatively large systematic plus random errors ($\sim$\,0.2\,--\,0.4\,dex) of  log\,$g$ estimates delivered by the LSP3, as shown by a recent comparison of LSP3 and asteroseismic  log\,$g$ values for common objects in the LAMOST--$Kepler$ fields (Ren et al. 2015, in preparation).    
As we shall show later (\S{2.2.2};  Fig.\,4) there are also significant deviations between $T_{\rm eff}$--log\,$g$ values yielded by the LSP3 and those predicted by the  PARSEC stellar isochrones (Bressan et al. 2012) of LSS-GAC DR2 red giant stars for a typical  metallicity [Fe/H]\,=\,$-0.3$.  
Therefore,  for selecting a clean RC sample from the LSS-GAC data set, it is necessary to improve the accuracy of LSP3 $\log\,g$ estimates..

\subsubsection{Re-determinations of LSP3 log\,$g$ estimates with the  KPCA  algorithm}
%---> re-determinations
Asteroseismology is very powerful for determination of stellar surface gravity log\,$g$, and is capable of yielding values much more accurate  than possible with the traditional  spectroscopic method.
Thanks to the $Kepler$ mission (Borucki et al. 2010), asteroseismic values of log\,$g$ of tens of thousands stars are now available, with a typical error of $\sim$\,0.02\,dex (Huber et al. 2014).
This  number of stars is however still orders of magnitude smaller than those already been targeted by  large-scale spectroscopic surveys, including LSS-GAC.
It will be thus very useful to calibrate and improve the accuracy of spectroscopic estimates of  log\,$g$ using the much more accurate asteroseismic values.
{Very recently, Liu et al. (2014) improve the accuracy (by a factor of two) of $\log\,g$ estimates of  LAMOST Stellar Parameter Pipeline (LASP; Wu et al. 2011; Luo et al. 2015) using a  support vector regression model  trained by asteroseismic \logg\, measurements of thousands of giants from the LAMOST--$Kepler$ fields.}
Here, a similar effort is presented  for the LSP3 \logg\,  estimates .
%Such an effort is presented here.
We re-determine and improve the accuracy of  LSP3 estimates of log\,$g$ based on a Kernel Principal Component Analysis (KPCA; Sch\"olkopf et al. 1998) method trained with thousands of giants in the LAMOST--$Kepler$ fields with accurate  asteroseismic log\,$g$ measurements.
%{ Very recently, a similar work by Liu et al. (2014) improve the accuracy of the LAMOST stellar parameter pipeline (LASP; Wu et al. 2011; Luo et al. 2015) estimates of \logg\, using support vector machine trained also with asteroseismic \logg\, measurements from $Kepler$.}

% KPCA and action
The Principal Component Analysis (PCA) is a classical statistical method that can  convert the observations (e.g. spectra) into a set of linearly uncorrelated orthogonal variables or principal components (PCs).
The method has widely been used to estimate stellar atmospheric parameters (e.g. Singh et al. 1998;  Rees et al. 2000; {Re Fiorentin et al. 2007}).  
The KPCA remains the concept of PCA but can be extended to nonlinear feature extraction using the kernel techniques.
Considering the nonlinear dependence of spectral features on the log\,$g$, the KPCA is more suitable  than the standard PCA method for the case of  log\,$g$ estimation.  

To apply the KPCA method to the $\log\,g$ determinations for a whole spectroscopic sample, it is necessary to construct a tight relation between the LAMOST spectra (at the present, only the blue-arm spectra\footnote{In principle,  red-arm spectra should also be included given that the stars of concern here are mainly red giants that are bright in the red and thus generally have high spectral SNRs. However, considering the sky subtraction of the current LAMOST 2D pipeline (Luo et al. 2015)  which still suffers from some problems in the red and the fact that most stars typically have few spectral features in the red, we have opted to  consider the blue-arm spectra in the current study. In the future, with improved sky subtraction, we expect including the red-arm spectra can help improve the results.}  between  3930\,$\leq$\,$\lambda$\,$\leq$\,5500\, \AA are used) and the corresponding asteroseismic values of  log\,$g$ with training data.
To build up a training sample, we have cross matched our sub-sample of about the 50,\,000 stars in LAMOST--$Kepler$ fields with the currently available largest asteroseismic log\,$g$ (hereafter $\log\,g_{\rm AST}$) sample ($\sim$\,16,\,000 stars) from Huber et al. (2014).
 In total 3562 common sources of spectral S/N (4650\,\AA)\,$\geq\,10$ are identified.
 { For self-consistency, we have used the LSP3 estimates of $T_{\rm eff}$ and  the observed frequencies of maximum power ($\nu_{\rm max}$) from Huber et al. (2014) to re-estimate  values of \logg\, for the 3562 common sources, using the asteroseismic scaling relation $\nu_{\rm max}\,\propto\,g\,T_{\rm eff}^{-0.5}$ (Brown et al. 1991; Belkacem et al. 2011).}
We then trim this sample of the common sources requiring that  LSP3 log\,$g$\,$\leq$\,$3.8$, $3500$\,$\leq$\,$T_{\rm eff}$\,$\leq$\,$6000$\,K and {[Fe/H]\,$\geq$\,$-1.0$},  for the following three reasons: 
1) The current study concerns mainly the estimation of  log\,$g$ for red giant stars including RGB and RC stars; 
2) Currently only a small number of dwarfs have asteroseismic log\,$g$ determinations; and finally,
3) The surface gravity  of metal-poor stars with {[Fe/H]$\,<\,-1.0$} are poorly determined currently by astroseismic scaling relation (Epstein et al. 2014).  
The trimming leaves 3267 sources. 
The 3267 sources are then further divided into two groups: a training sample and a control sample.
For the training sample, we select 1355 stars with spectral S/N (4650\,\AA)\,$\geq$\,$50$ to minimize the effects of  spectral noises and of errors of LSP3 stellar atmospheric parameters $T_{\rm eff}$ and [Fe/H] in the training process.
The remaining 1912 stars constitute the control sample, which will be used to test the accuracy of log\,$g$ determined with KPCA  algorithm.

\begin{figure*}
\centering
\includegraphics[scale=0.45,angle=0]{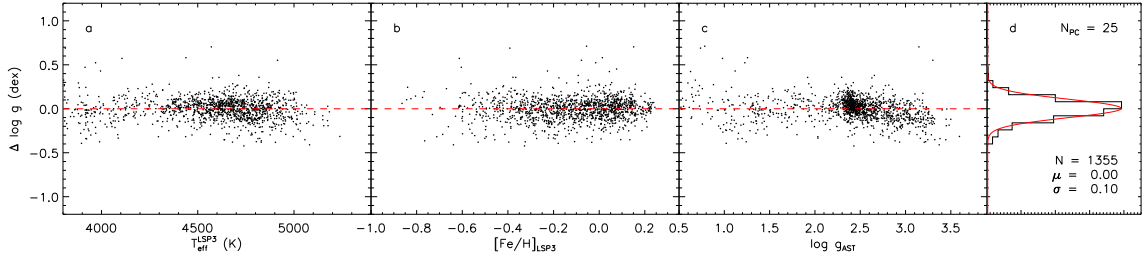}

\caption{Distributions of values of residual, log\,$g_{\rm KPCA}-$log\,$g_{\rm AST}$, {of the training sample} as a function of LSP3 $T_{\rm eff}$ and [Fe/H] (Panels a and b) and of asteroseismic log\,$g$ (Panel c).
Panel\,(d) shows a histogram distribution of the residuals (black line).
Also overplotted in red is a Gaussian fit to the distribution.
The mean $\mu=0.00$\,dex and dispersion $\sigma=0.10$\,dex of the fit, as well as the number of stars used, $N=1355$, are marked. 
For the current KPCA analysis, $N_{\rm PC} = 25$ is assumed. See text for detail.}
\end{figure*}

\begin{figure}
\centering
\includegraphics[scale=0.6,angle=0]{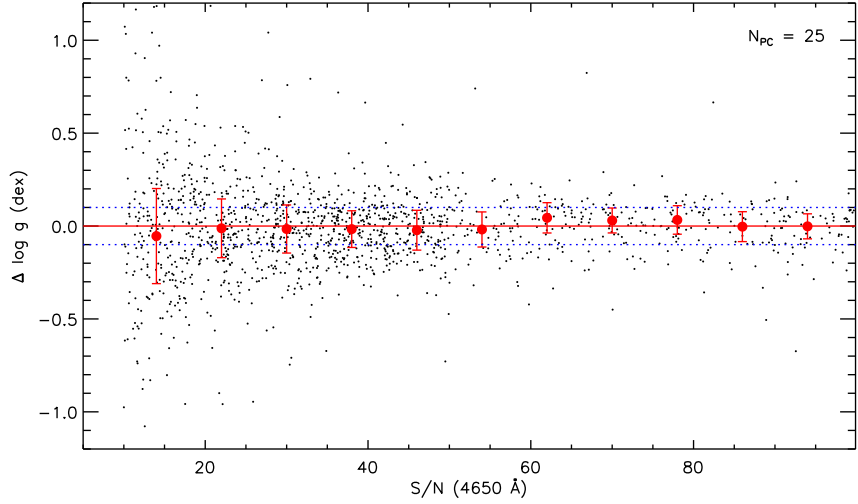}

\caption{Differences of  KPCA log g estimates and asteroseismic values for  the control sample, plotted as a function of spectral  S/N (4650\,\AA). 
The blue dashed lines indicate differences $\Delta$\,log\,$g$\,$=$\,$\pm$\,0.1\,dex.
The means and standard deviations of the differences in the individual S/N (4650\,\AA) bins (binsize\,$=$\,8 ) are oveplotted with dots and error bars, respectively.
 Again, $N_{\rm PC} = 25$ is assumed.}
\end{figure}

\begin{figure}
\centering
\includegraphics[scale=0.6,angle=0]{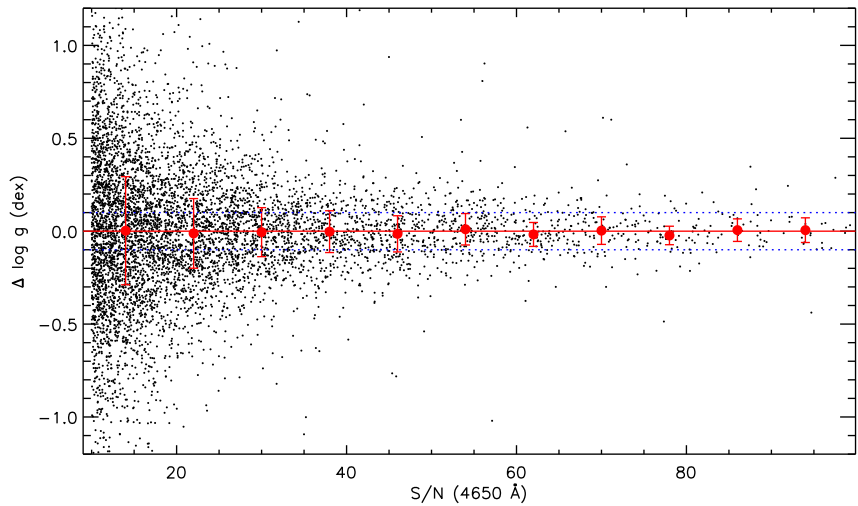}

\caption{Differences of  KPCA log g estimates yielded by  duplicate observations of similar S/N (4650\,\AA)  plotted  as a function of  S/N (4650\,\AA).
The blue dash lines indicate differences  $\Delta$\,log\,$g$\,$=$\,$\pm$\,0.1\,dex.
The means and standard deviations  of the differences  in the individual S/N (4650\,\AA) bin (binsize\,$=$\,8) are oveplotted as red  dots and error bars, respectively.
 Again, $N_{\rm PC} = 25$.}
\end{figure}

\begin{figure*}
\centering
\includegraphics[scale=0.7,angle=0]{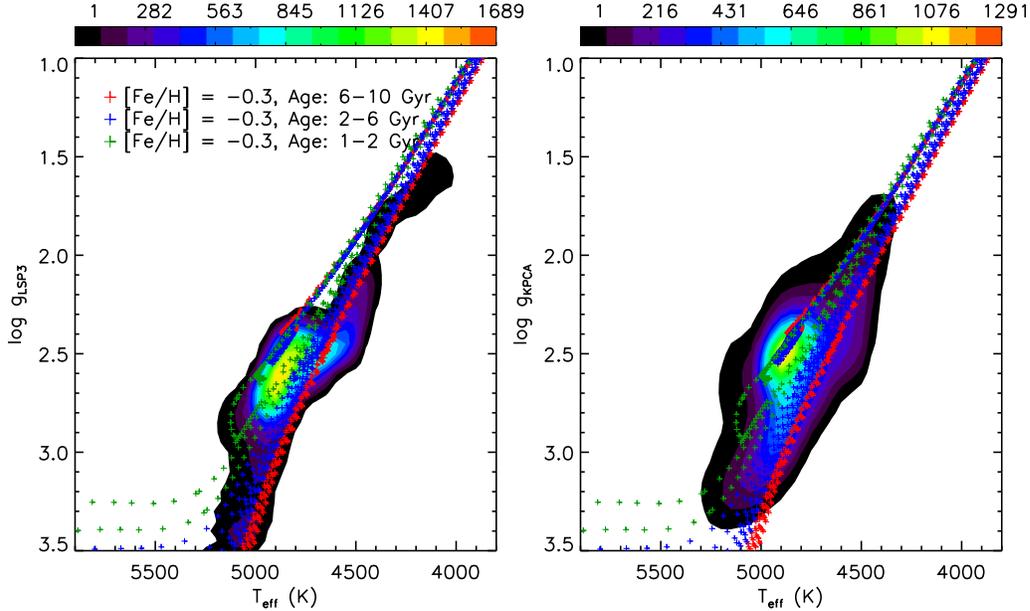}
\caption{Pseudo-color HR diagrams of red giant stars from the LSS-GAC DR2.
Only stars of LSP3 metallicity [Fe/H] between $-0.4$ and $-0.2$ (where the metallicity distribution of our red giant sample peaks) are included.
                   Effective temperatures are those given by the LSP3.
                   In the left panel, log\,$g$ values are from the LSP3 , whereas those in the right panel are newly derived with the KPCA method.
                  Also overplotted are the PARSEC stellar isochrones for [Fe/H]\,$= -0.3$.
                  Different colors represent different stellar ages (1--2 Gyr, green crosses; 2--6 Gyr, blue crosses; 6--10 Gyr, red crosses).}
\end{figure*}

\begin{figure*}
\centering
\includegraphics[scale=0.35,angle=0]{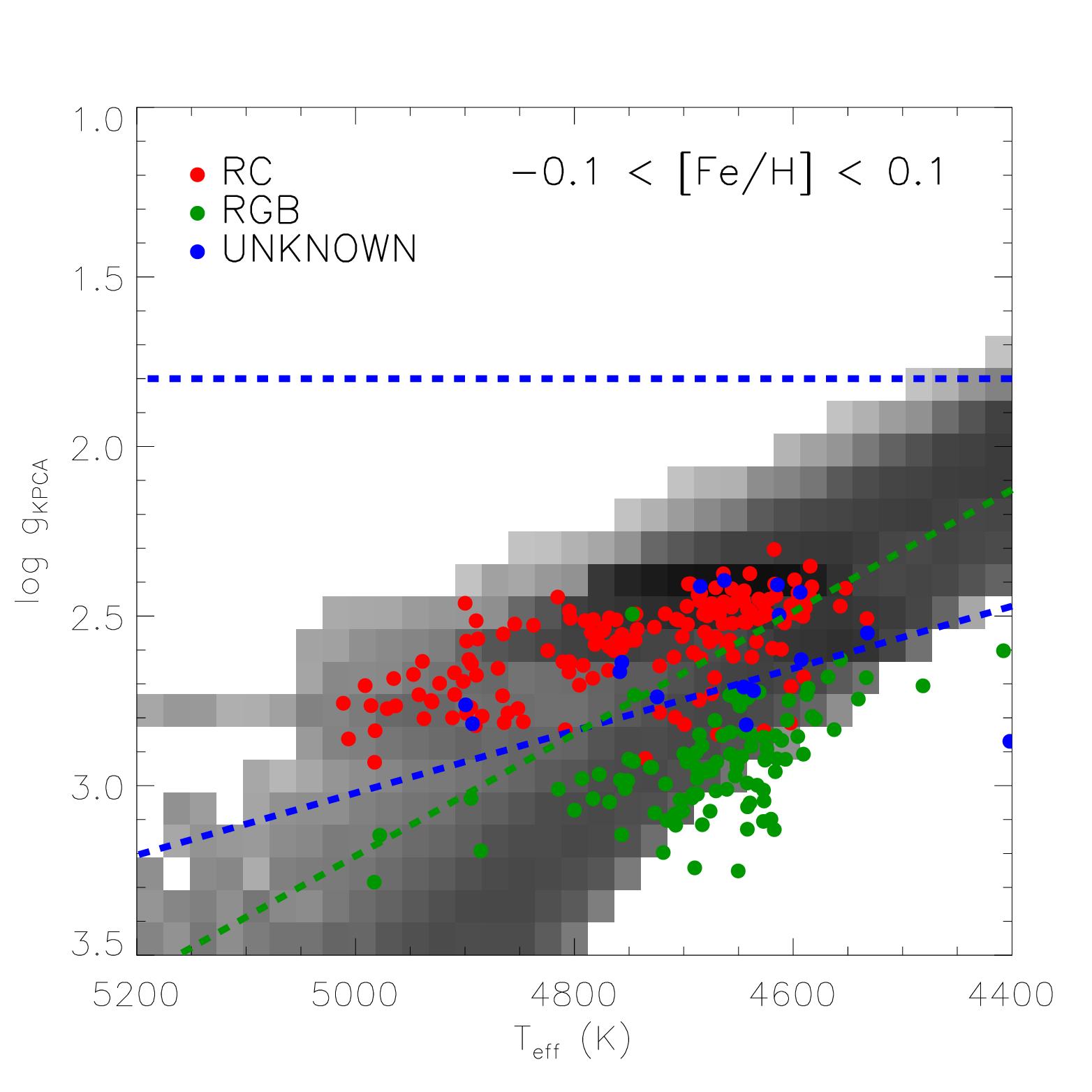}
\includegraphics[scale=0.35,angle=0]{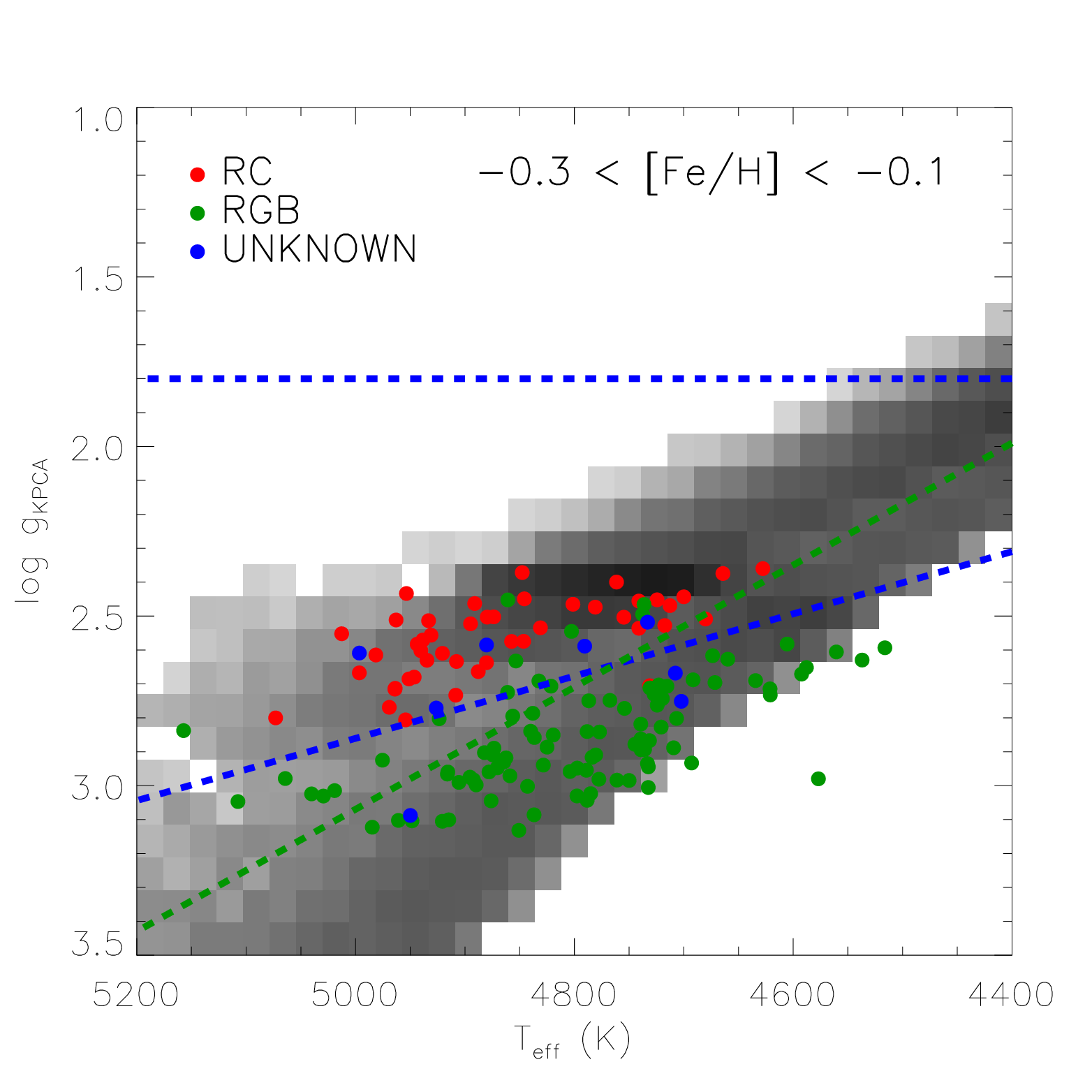}
\caption{Distribution of stars on the $T_{\rm eff}$\,--\,\logg\, HR diagram  predicted by the PARSEC stellar evolution model with assumptions as described  in the text (grey scale), compared to that of the LE sample stars (colored points), for two metallicity bins as marked.
Values of $T_{\rm eff}$ and [Fe/H] of the LE sample stars are from LSP3 and those of $\log\,g$ from the KPCA method.
Red, magenta and blue points represent stars of  RC, RGB and unknown evolutionary stages, as classified by Stello et al. (2013), respectively.
The blue dashed lines represent the cuts that separate RC stars and  the less luminous RGB stars as given in  Eqs.\,(1) and (2).
The green dashed lines give  the cuts adopted  by B14. }
\end{figure*}

To extract information of the stellar surface gravity from the LAMOST spectra, we assume a  KPCA model (with a Gaussian kernel) of 25 Principle Components (PCs), $N_{\rm PC} =25$, when constructing the relation between log\,$g_{\rm AST}$ and the LAMOST blue-arm spectrum with the training sample. 
{$N_{\rm PC} =25$ is chosen in the KPCA model for a tradeoff between reducing the training residual and retaining a relative high accuracy of \logg\, estimate for LAMOST spectra at low spectral  S/N's (see Appendix A).}
In the training process, $T_{\rm eff}$ and [Fe/H] are fixed to the value yielded by the LSP3.
The model  fits the data well, yielding residuals of only $\sim$\,0.1\,dex, as shown in Fig.\,1. 
In addition, the Figure  shows that the residuals exhibit no obvious systematic trend as the LSP3 $T_{\rm eff}$ and [Fe/H] vary.
{ But we caution that the residuals show a weak trend as the asteroseismic \logg\, varies from 2.5 to 3.5.
The reason is unclear but the trend do not affect the selection of RC stars (see Fig.\,5) .}
To test the accuracy of KPCA log\,$g$  values, we apply the relation yielded by the training sample to the control sample. 
The results are  shown in Fig.\,2.
The differences between KPCA  and asteroseismic log\,$g$ are less than 0.15\,dex for spectral S/N (4650\,\AA)\,$>18$ and less than 0.10\,dex for S/N\,(4650\,\AA)\,$>30$.
 Thus the accuracy of KPCA log\,$g$ estimates are about a factor of two  better than the LSP3  values.
The test shows that there is no obvious systematics over the whole parameter space (atmospheric parameters as well as S/N ratios) spanned by the control sample.
 
 %Results
We now apply the above model to the LAMOST blue-arm spectra and re-determine log\,$g$ values for all LSS-GAC DR2 red giants\footnote{Red giants are selected from the whole LSS-GAC DR2 sample using the same set of stellar atmospheric parameter cuts as for  the training sample, i.e. log\,$g$\,$\leq$\,$3.8$, $3500$\,$\leq$\,$T_{\rm eff}$\,$\leq$\,$6000$\,K and [Fe/H]\,$\geq$\,$-1.5$ .} using the relation constructed from  the training sample.
This yields  newly derived KPCA log\,$g$ values  for over 0.35\,million LSS-GAC DR2 red giants\footnote{This includes targets from the LSS-GAC as well as  stars from the spheroid part of the LEGUE and the LAMOST--$Kepler$ fields.} with spectral S/N\,(4650\,\AA)\,$>$\,$10$. 
As another test of the accuracy of newly derived KPCA \logg\, estimates, we plot the differences of KPCA $\log\,g$ values yielded by  $\sim$\,$6500$ pairs of  duplicate observations of similar S/N\,(4650\,\AA) as a function of S/N.
The results are quite similar to those found for the control sample. 
Again, this indicates that our newly derived KPCA\,\logg\,  values are about a factor of 2 more accurate than those of  LSP3.
%With this check, we give a look table for the accuracy of \logg\, at a given S/N\,(4650\,\AA).
We also find that the HR diagram constructed using the LSP3 $T_{\rm eff}$ and the newly derived KPCA \logg\,  for our sample stars is in excellent agreement with the prediction of the  PARSEC stellar evolution model, as shown in Fig.\,4.
As mentioned before, the position of  RC clump in the original HR diagram constructed  with LSP3 $T_{\rm eff}$ {\em and}  \logg\, deviates significantly from  that predicted by the PARSEC model.

\begin{figure}
\centering
\includegraphics[scale=0.70,angle=0]{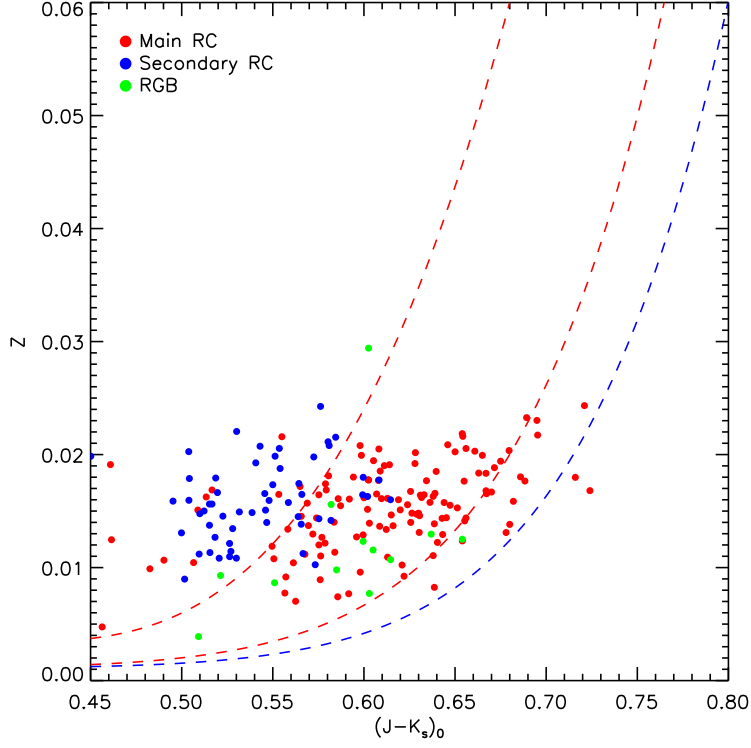}
\caption{Distribution of RC-like stars in the color $(J-K_{\rm s})_{0}$ Z plane, using 2MASS photometric measurements and LSP3 metallicities.
Dots in red, blue and green  represent main RC, secondary RC and RGB stars, respectively.
The red dashed lines are developed by B14 to eliminate the contamination of secondary RC stars.
The blue dashed line is our revised cut  developed in the current study  in replacement of the lower red dashed line.}
\end{figure}

\subsubsection{Selection of RC stars}
%RC selections
As described above, the newly derived KPCA \logg\, values, with much improved accuracy (both in random and systematic), have much reduced the differences between the observed and predicted position and morphology of the   RC clump on the HR diagram, thus enabling us to select  a clean RC sample following the method proposed by B14.
Firstly, to obtain a set of  cuts that distinguish RC and RGB stars on the $T_{\rm eff}$\,--\,\logg\, diagram, we cross-identify red giants from the LAMOST-$Kepler$ fields with those from Stello et al. (2013), in which the evolutionary stages of stars have been  classified based on the frequencies and spacings of gravity-mode measured from the $Kepler$ light curves. 
This leads to  common  600 stars with LAMOST spectral S/N(4650\,{\AA}) $\geq 40$  (hereafter the LAMOST-EVOLUTION sample, the LE sample for short).
The S/N cut is to ensure that the atmospheric stellar parameters are accurate enough, especially the KPCA \logg\, values (better than 0.1\,dex under this cut).
The LE sample stars are mostly classified by Stello et al. (2013)  as either  `RC' or `RGB', only a few  as `UNKNOWN'. 
Unlike B14, who use cuts defined with the asteroseismic \logg\, values, we opt to define the cuts  using the LE sample,  relying directly  on log\,$g$ values determined with the KPCA method, together with $T_{\rm eff}$ and [Fe/H] yielded by the LSP3.
With this change, there is no need to introduce an additional cut that correct for the systematic difference between the asteroseismic log\,$g$ [e.g. the additional cut given by Eq.\,(9) of B14], nor the need to account for the possible systematic differences between the PARSEC and spectroscopic values of the other atmospheric parameters ($T_{\rm eff}$ and [Fe/H]).
The cuts defined here are therefore more straightforward and appropriate for real observed data.
Fig.\,5 compares the distribution on the $T_{\rm eff}$ -- $\log\,g$ HR diagram of the LE sample stars and that predicted by the PARSEC stellar isochrones for two  metallicity bins.
The PARSEC distribution is generated assuming, similar to B14, a lognormal Chabrier (2001) for initial mass function (IMF), a constant SFH with ages less than 10 Gyr and metallicity distribution functions (MDFs) that match with that of  the LE sample stars in the two metallicity bins.
As  Fig.\,5 shows, the distribution of  RC-like stars,  with \logg\, values  roughly between 2.5 and 2.9, and that of  RGB stars, as classified by Stello et al. (2013), is { roughly in agreement with the predicted distribution from the PARSEC stellar evolution model except for some small systematic differences between the LSP3 and PARSEC values of $T_{\rm eff}$}, and is well separated from that of RGBs as well.
Finally, ignoring those  stars classified as  `UNKNOWN' evolutionary stage, we derive the following  cuts as given by the following two equations  that maximize the product of completeness $f_{\rm comp}$ (${\equiv N_{\rm RC}^{\rm Sel}/N_{\rm RC}^{\rm All}}\,\sim\,96\,\%$) and purity $f_{\rm purity}$ ($\equiv 1- N_{\rm RGB}^{\rm Sel}/N_{\rm All}^{\rm Sel}\,\sim\,95\,\%$),
\begin{equation}
\centering
1.8 \leq {\rm log}\,g \leq \,0.0009\,{\rm dex\,K}^{-1}\,\{T_{\rm eff}-T_{\rm eff}^{\rm Ref}{\rm ([Fe/H])\}}+2.5\,, 
\end{equation}
where
\begin{equation}
\centering
T_{\rm eff}^{\rm Ref}{\rm ([Fe/H])} = -876.8\,{\rm K}\,{\rm dex}{^{-1}}\,{\rm [Fe/H]} +4431\,{\rm K}\,.
\end{equation}
${N_{\rm RC}^{\rm All}}$ represents the total number of  real RC stars,  and ${N_{\rm RC}^{\rm Sel}}$ and ${N_{\rm RGB}^{\rm Sel}}$  the numbers of RC and RGB stars that pass the cuts of Eqs.\, (1) and (2), respectively.
${\rm {\it N}_{All}^{Sel}}$ is the sum of 
${\rm  {\it N}_{RC}^{Sel}}$ and ${\rm {\it N}_{RGB}^{Sel}}$, that represents the total number of stars passing the cuts. 

Fig.\,5 shows that the cuts developed by B14  (green lines  in the Figure) is not fully suitable for our data because their cuts reject some real RC stars in the lower temperature range but include some RGB stars as RC stars in the higher temperature range.
We have also plotted Fig.\,5 using the asteroseismic values of  \logg\, from Huber et al. (2014) for the LE sample stars  do not find any significant difference from what shown here.
This indicates the difference between our current  cuts and those of B14 can not be caused by  the usage of  KPCA \logg. 
{ This difference in cuts is possibly caused by the systematic differences in values of  $T_{\rm eff}$ as yielded by the LSP3  and given by PARSEC, as mentioned earlier.}
%the coarse sampling of the lower temperature RC stars of the calibrated sample used by B14 that leads their cuts to miss the RC stars in the lower temperature range.
%The systematic differences in $T_{\rm eff}$ between LSP3  and APOGEE also can not be ruled out.

To examine the performance of our cuts for different errors of \logg\, measurements, we use a Monte Carlo method to simulate the selection  process under different errors of \logg\, and derive the corresponding values of $f_{\rm comp}$ and $f_{\rm purity}$.
Here we assume that  the uncertainties of the LE sample stars have a minimum value of 0.1\,dex (almost the best accuracy achievable using  the KPCA method), based on the test results presented in Section 2.2.2  using the control sample and duplicate observations.
Values of $f_{\rm comp}$ and $f_{\rm purity}$ for larger errors are then derived by assigning larger errors randomly to \logg\, values of the LE stars and repeating the selection with Eqs.\,(1) and (2) by 10,\,000 times.
The results are presented in  Table\,1.  
In Table\,1, we have also listed the spectral S/N(4650\,{\AA}) ratios required by the KPCA method in order to achieve the listed uncertainties of $\log\,g$ measurements. 
With the cuts developed in the current work, the amount of contamination from RGB stars is  twice less than that for selection using a ``small box''  such as that adopted by  Williams et al. (2013), even for \logg\, measurement uncertainties as large as  0.3\,dex , an accuracy achievable with the KPCA method even for a spectral  S/N (4650\,\AA)\, ratio as low as\,10.

\begin{table*}
\centering
\caption{Values of  purity and completeness of selection of RC-like stars  under different uncertainties  errors of \logg\, measurements for the cuts given by Eqs.\, (1) and (2).}
%\begin{center}
\begin{threeparttable}
\begin{tabular}{cccccc}
\hline
$\sigma_{{\rm log}\,g}$\,(dex)&0.1&0.15&0.20&0.25&0.30\\
\hline
S/N$^a$&$>\,40$&20&18&15&10\\
$f_{\rm purity}$&$94\pm1$\%&$89\pm1$\%&$83\pm2$\%&$78\pm2$\%&$74\pm2$\% \\
$f_{\rm comp}$&$96\pm1$\%&$90\pm2$\%&$83\pm2$\%&$78\pm2$\%&$74\pm3$\% \\
 \hline
\end{tabular}
%\end{center}
\begin{tablenotes}\small
\item[$a$]S/N ratios required by the KPCA method to achieve the corresponding uncertainties of $\log\,g$ measurements listed in the first row of the Table.  
\end{tablenotes}
\end{threeparttable}
\end{table*}

As shown in Fig.\,5, the region where RC-like stars fall  show two distinct features: a clump of stars of almost constant \logg\,$=$\,$2.5$ and $4600$\,$\leq$\,$T_{\rm eff}$\,$\leq$\,$4850$\,K (i.e. the main RC stars) and a skew tail of stars  with $2.5$\,$\leq$\,\logg\,$\leq$\,$2.9$ and $4800$\,$\leq$\,$T_{\rm eff}$\,$\leq$\,$5000$\,K (i.e. the secondary RC stars of fainter luminosities).
The secondary RC stars are not standard candles and their luminosities are a function of mass (age).
We need to remove them from our RC-like star sample to ensure that the distances of the sample stars can be determined precisely  and simply.

To remove secondary RC stars and low surface gravity  RGB stars, we adopt the color $(J-K_{\rm s})_{0}$\,--\,metallicity $Z$\footnote{Z is converted from [Fe/H] using the equation given by Bertelli et al. (1994), assuming $Z_\odot = 0.017$.} cuts as developed by B14 using the PARSEC stellar evolution model.
The cuts are aimed to select  stars that have a nearly constant absolute magnitude with 1$\sigma$ dispersion $\leq$\,$0.1$\,mag and eliminate those non-standard candles, including secondary RC, RGB and AGB stars.
For a more detailed description, please refer to the Section 2.2 of B14.
Since the cuts are purely based on stellar evolution model, empirical examination is required to eliminate the potential discrepancies, if any, between the model and observations. 
To test the cuts, we again use the LE sample.
For  consistency, we first apply the cuts  of B14 on the $T_{\rm eff}$\,--\,\logg\, diagram  to the LE sample to select RC-like stars using  the asteroseismic \logg\, values.
After corrected for the interstellar extinction as estimated with the `star pair' method (see Yuan et al. 2015), the distribution of selected RC-like stars in the  $(J-K_{\rm s})_{0}$\,--\,$Z$  plane is shown in Fig.\,6. 
The  photometric magnitudes of the stars are taken  from 2MASS (Skrutskie et al. 2006) and their  metallicities  from the LSP3.
By definition, secondary RC stars have masses\,$\geq\,2\,M_{\rm \odot}$ and main RC stars have masses\,$<\,2\,M_{\rm \odot}$.
Again with masses estimated also by Stello et al. (2013), we divide the RC-like stars into two groups: main and secondary RC stars.
As Fig.\,6 shows, the contaminations of RGB stars, as expected, is negligible but that  of secondary RC stars is significant.
Luckily, compared to  main RC stars,  secondary RC stars belong to a young and metal-rich population of large masses
As a consequence,  they occupy the bluer and more metal-rich parts of color\,--\,metallicity diagram. 
Main and secondary RC stars can be well separated using a cut given by the following equation as proposed by B14 (the upper red dashed line in Fig.\,6), 
\begin{equation}
\centering
Z\,<\,2.58[(J-K_{\rm s})_{0}-0.400]^{3}+0.0034\,.
\end{equation}

However, Fig.\,6 also shows that the lower cut of B14 (the lower red dash line in Fig.\,6), designed to reject high surface gravity  RGB stars, seems to also eliminate a considerable fraction of real main RC stars.
This is possibly caused by discrepancies between model and observations.
To correct for this, we adjust the lower cut of B14 to (the blue dashed line in Fig.\,6), 
\begin{equation}
\centering
Z\,>\,1.21[(J-K_{\rm s})_{0}-0.085]^{9}+0.0011\,.
\end{equation}

\subsubsection{The RC sample}
%the sample
With $T_{\rm eff}$, [Fe/H] estimated with the LSP3, \logg\, derived with the KPCA method and $(J-K_{\rm s})_{0}$ calculated from 2MASS photomtry\footnote{Only stars with  {\it ph\_qual} flagged as `A' in both $J$ and $K_{\rm s}$ bands are included.} after corrected for extinction as estimated with the `star pair' method (Yuan et al. 2015), we apply the cuts of Eqs.\,(1)\,--\,(4)  to the LSS-GAC DR2 red giant sample and obtain a clean RC sample of over 0.11 million stars of S/N\,(4650\,\AA)\,$\geq\,10$.
The distances of these RC stars are derived using the recent calibration, $M_{K_{\rm s}}\,=\,-1.61$\,mag, for nearby RC sample (Laney et al. 2012)\footnote{B14 assign absolute magnitudes to RC stars based on their positions in the $(J-K_{\rm s})_{0}$\,--\,$Z$ plane. 
However, we find that the absolute magnitudes of RC stars predicted by the stellar evolution model are almost constant except for those super metal-rich ones ($Z$\,$\geq$\,0.04, i.e. [Fe/H]\,$\geq$\,0.4\,--\,0.5\,dex; see Fig.\,3 of B14), which are very rare in our RC sample.}.
Since the intrinsic scatter of absolute magnitudes of RC stars  is estimated to be within 0.1\,mag, the distances derived are expected to have uncertainties no more than 5\,--\,10\,\%, given  a typical photometric error of $\sim\,$0.05\,mag in $K_{\rm s}$-band and an extinction error of $\sim\,$0.04\,mag in $E(B-V)$ (see Yuan et al. 2015).
Proper motions of  the sample stars are taken  from the UCAC4 (Zacharias et al. 2013) and PPMXL (Roester et al. 2010) catalogue.s
%the potential
Given the high distance precision and large spatial coverage (Fig.\, 7) of the current RC star sample, it is not only useful to address  the problem of metallicity gradients of the Galactic disk of concern here,  but can also tackle a variety of problems with regard to the Galactic structures  and  dynamics.   
The sample will be publicly available at website:  (\url{http://162.105.156.249/site/RC\_Sample}). 

%For abundance:
In order to obtain accurate metallicity gradients of the Galactic disk(s) of concern here, we restrict the current analysis to a subset of the RC sample stars  with S/N (4650\,\AA)\,$>15$, to ensure a sufficiently high accuracy of [Fe/H] determinations, and  to eliminate possible contaminations  from halo stars of with [Fe/H]\,$\leq$\,$-1.0$\,dex (also the lowest applicant range in [Fe/H] of KPCA method in \logg\, estimates).
Finally, a total of 74,\,280 unique RC stars are selected.
The number density distributions of this adopted sample of  RC stars in the $X$\,--\,$Y$ and $X$\,--\,$Z$ Galactic planes are presented in Fig.\,7.
Here $X$, $Y$ and $Z$ are coordinates of a right-handed Cartesian coordinate system with an origin at the Galactic center. 
The $X$-axis passes through the Sun and points towards the Galactic Center, the $Y$-axis is in
the direction of Galactic rotation and the $Z$-axis points towards the North Galactic Pole. The Sun is located at ($X$, $Y$, $Z$) = ($-R_{\rm 0}$, 0, 0), where $R_{\rm 0}$ = 8 kpc.
As Fig.\,7 shows, our RC sample covers a volume of  $-16$\,$\leq$\,$X$\,$\leq$\,$-6$\,kpc, $-4$\,$\leq$\,$Y$\,$\leq$\,$5$\,kpc and $-3$\,$\leq$\,Z\,$\leq$\,$3$\,kpc.

\begin{figure*}
\centering
\includegraphics[scale=0.28,angle=0]{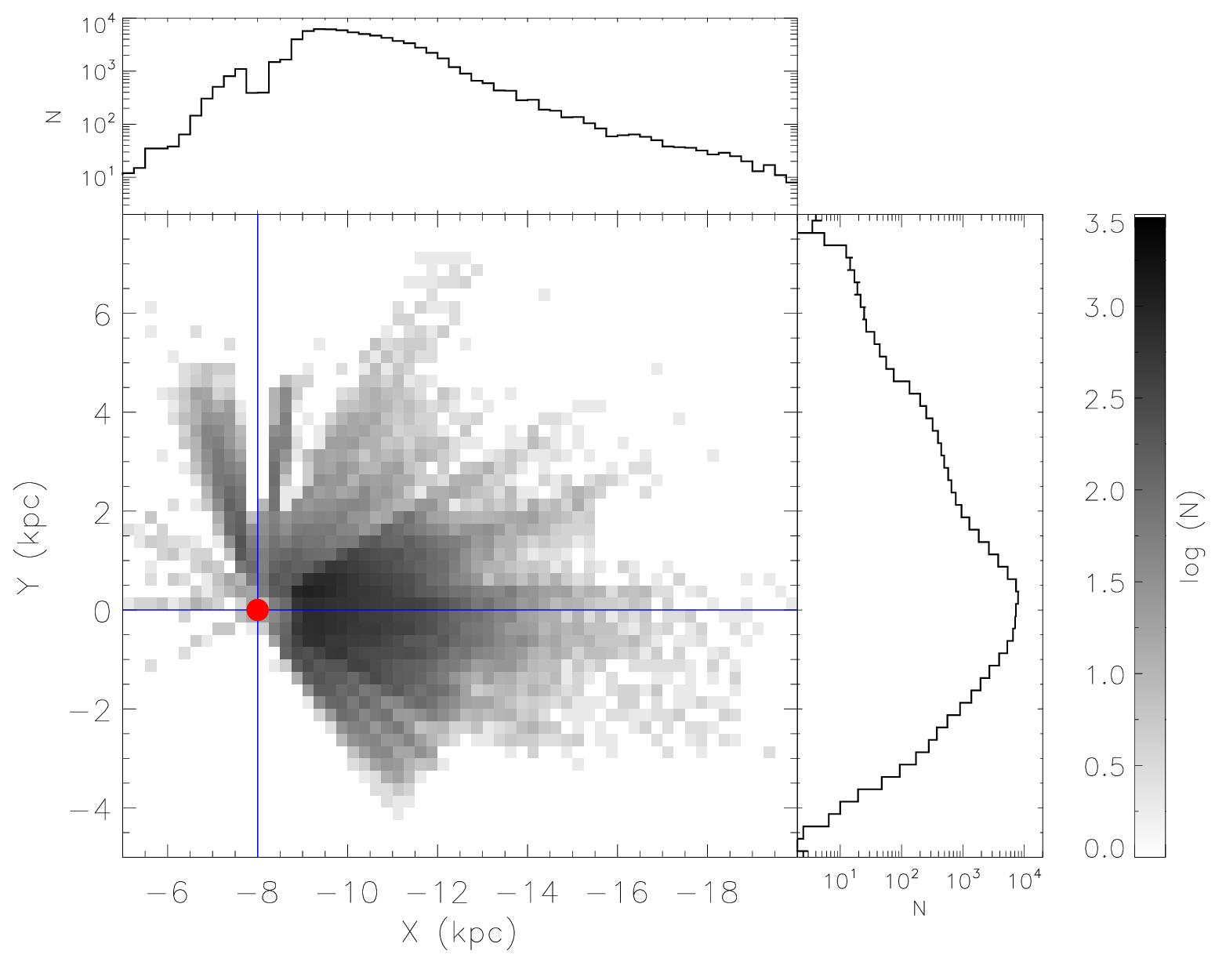}
\includegraphics[scale=0.28,angle=0]{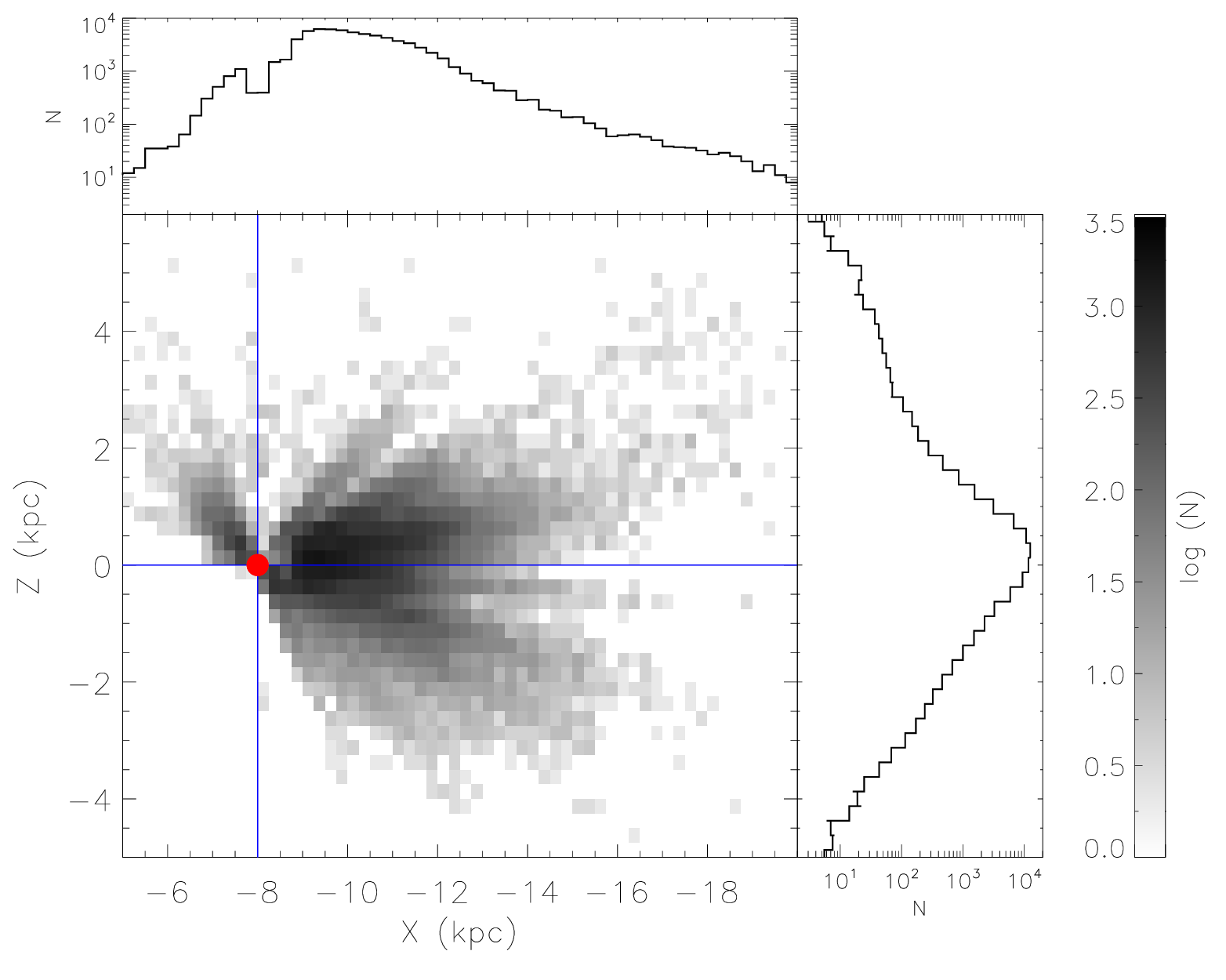}
\caption{Greys cale number density distribution of RC sample stars in the $X$\,--\,$Y$ (left panel) and $X$\,--\,$Z$ (right panel) planes. 
The Sun is located at ($X$, $Y$, $Z$) = ($-8.0$, 0.0, 0.0) kpc.
The stars are binned by 0.25$\times$0.25 kpc$^{2}$ in both diagrams. 
The densities are shown on a logarithmic scale.
Histogram distributions along the $X$, $Y$ and $Z$ axes are also plotted.}
\end{figure*}

%the selection bias
We note that  the current RC sample is largely  a magnitude limited sample.
Considering that the absolute magnitudes of RC stars are quite insensitive to both metallicity and age, we believe that any potential population effects that may be present in our sample are unlikely to significantly affect the metallicity gradients derived from this sample.
In addition, the cuts to define the RC sample in the current selection excludes stars of ages younger than 1\,Gyr (i.e. those relatively massive and young secondary RC stars) and halo stars of  metallicities poor than $-1.0$\,dex, the current sample is expected to be mainly composed of stars of intermediate-age stars with typical ages between 1.0 and 4.0 Gyr { if one assumes a flat SFH and a solar-neighborhood MDF (Casagrande et al. 2011; see Section\,5 of B14 for a more detailed discussion).}

\section{Results}
\subsection{Spatial distribution of mean metallicities}
The spatial distribution of mean metallicities of the RC sample in 0.40\,$\times$\,0.25\,kpc$^{-2}$ bins, with at least five stars in each bin, in the projected Galactocentric radius ($R_{\rm GC}$) and the distance from the  midplane ($|Z|$) is presented in Fig.\,8.
{ For each bin, the mean [Fe/H] is derived.}
Fig.\,8 shows several obvious features:

1) A gradient in both the radial and vertical direction is  evident, especially near the solar circle;

2) By the midplane, the radial  gradient near the solar circle is steeper than that in the outer disk (see Section 3.2);

3) The radial gradient flattens as one moves away from the  midplane (see Section 3.2);

4) The vertical gradient near the solar circle is steeper than that at large $R_{\rm GC}$ (see Section 3.3).
  
To describe those features of the spatial metallicity distribution in a quantitive manner, we measure the radial and vertical gradients separately  for the different regions of the disk in the following subsections.

\subsection{Radial metallicity gradients}

\begin{figure}
\centering
\includegraphics[scale=0.5,angle=0]{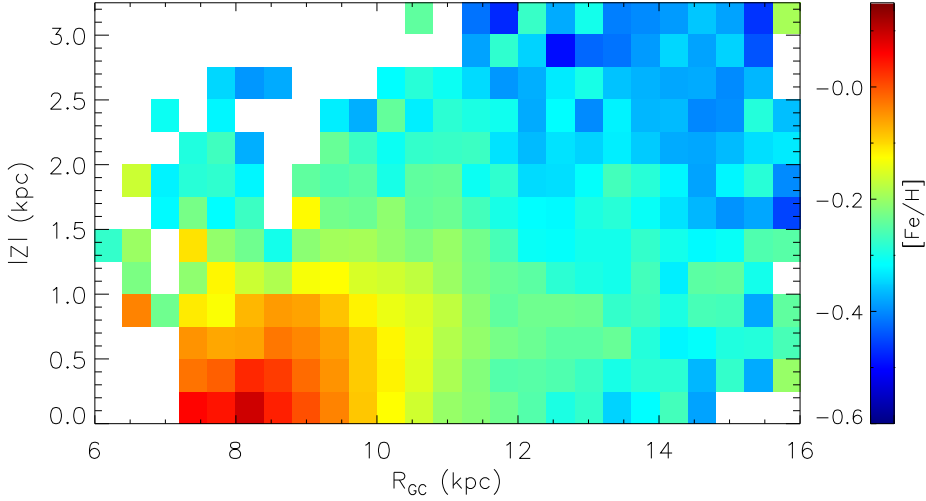}
\caption{Mean metallicity [Fe/H] as a function of $R_{\rm GC}$ and $|Z|$ deduced from  our RC sample, binned by  0.40\,$\times$\,0.25\,kpc$^{-2}$  in $R_{\rm GC}$ and $|Z|$, respectively.}
\end{figure}

%\subsubsection{Radial gradient as a function of $|Z|$}

\begin{figure*}
\centering
\includegraphics[scale=0.32,angle=0]{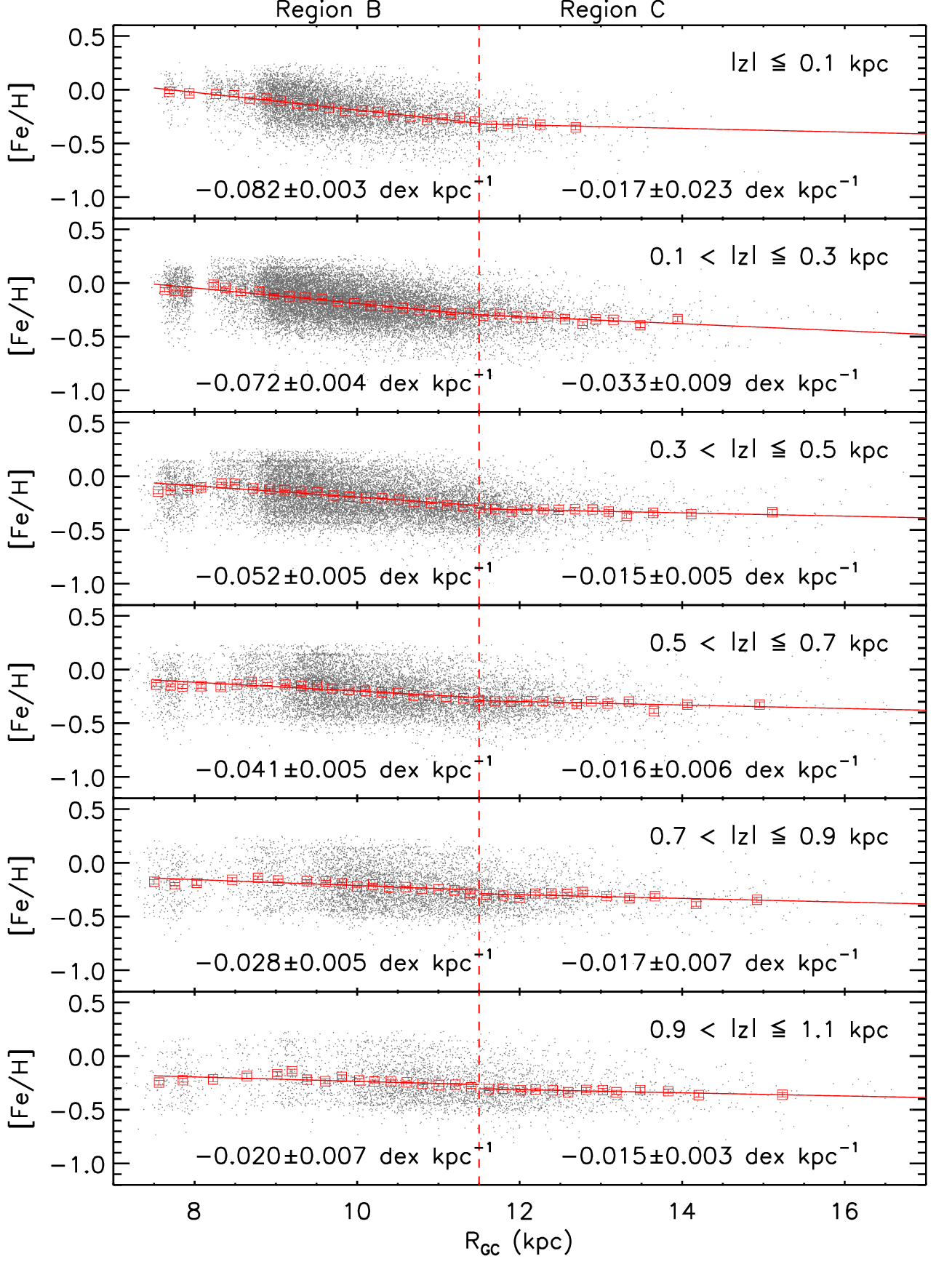}
\includegraphics[scale=0.32,angle=0]{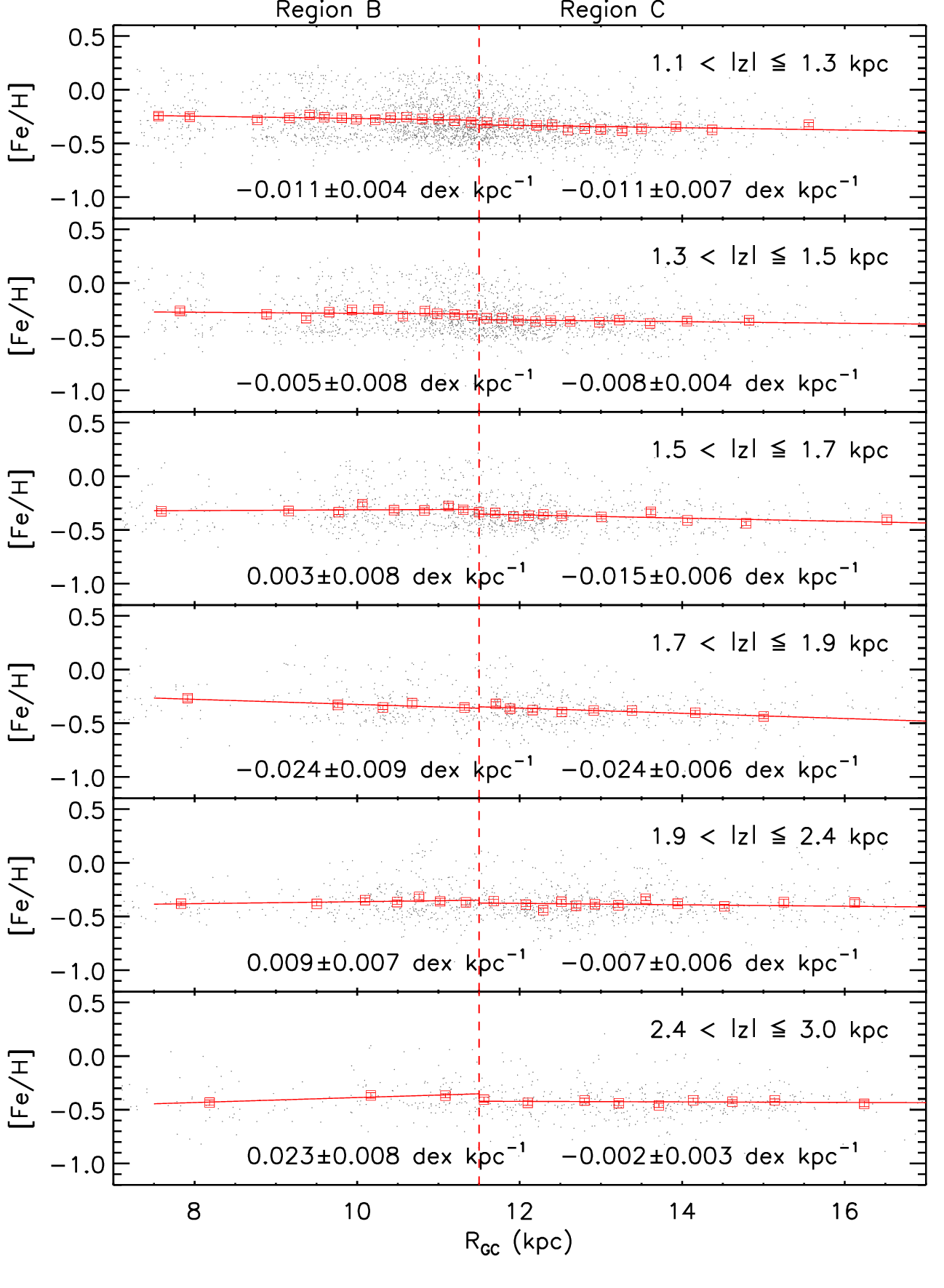}
\caption{Metallicity distribution as a function of $R_{\rm GC}$ in the individual disk height bins derived from our RC sample. 
The red dashed lines divide the disk into two parts: Region B near the solar circle and region C for the outer disk, as marked on top of the two columns of plots.
Red boxes are the mean metallicities in the individual radial annuli.
The metallicity  gradients of Regions B and C are fitted separately, each by a straight line. 
The results are  shown by red lines.
The gradient s and errors thus derived  are also marked near  the bottom of each panel.}
\end{figure*}

\begin{figure}
\centering
\includegraphics[scale=0.6,angle=0]{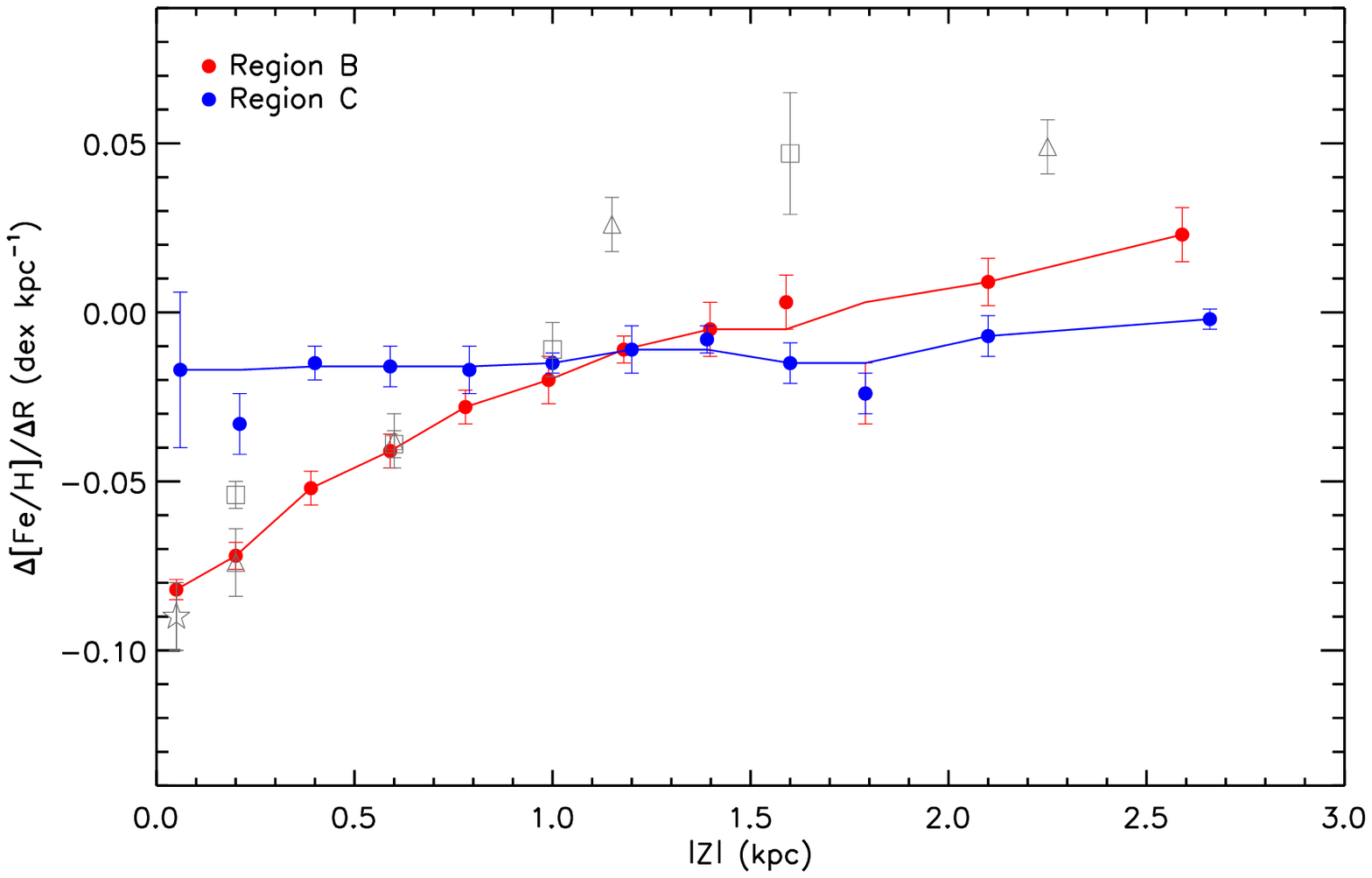}
\caption{Radial gradients, $\Delta$[Fe/H]/$\Delta R_{\rm GC}$, as a function of $|Z|$ yielded by the RC sample for disk Regions B (red dots) and C (blue dots).
The lines connecting the dots have been  smoothed over 3 adjacent points.
The symbols in grey represent measurements from the literature, all derived from tracers near the solar circle (i.e. Region\,B of this study).
The star is the measurement  by B14  using APOGEE RC stars,
triangles the measurements by Anders et al.(2014) using  APOGEE giants and
boxes the measurements by Boeche et al. (2014) using  RAVE RC stars.}
\end{figure}

The radial and vertical gradients, $\Delta$[Fe/H]\,/\,$\Delta R_{\rm GC}$ and  $\Delta$[Fe/H]\,/\,$\Delta Z$, are generally correlated with each other.
Fortunately, with a large, contiguous disk volume sampled by a large number of stars in hand  as in the current case, it is now possible to determine  the radial (vertical) gradient in thin vertical (radial) slice (annulus), and thus quantify the variations in both directions independently.

To derive the radial gradient from our RC sample, we divide the stars into bins of constant height, $\Delta |Z| =  0.2$\,kpc, except for the inner most bin of  $|Z|\,<0.1\,$kpc and the two outer most bins of $1.9\,<\,|Z|\,\leq\,2.4$\,kpc and $2.4\,<\,|Z|\,\leq\,3.0$, respectively.
For each bin, a straight line is adopted to fit the mean metallicities of the individual  radial annuli as a function of $R_{\rm GC}$.
The slope of the straight line gives the radial gradient.
To ensure a sufficient number of stars in all radial annuli, the binsize in radial direction  is allowed to vary. 
We require that the radial binsizes are no thinner than 0.2\,kpc and each radial annulus contains no less than 50 stars.

Motivated by the previous findings  that the metallicity variations in the radial direction across the entire range of the disk may not be captured by a single slope, we, inspired by Magrini et al. (2009), divide the whole disk  into three parts: Region A of  $3\,\leq\,R_{\rm GC}\,\leq\,6$\,kpc,  Region\,B of $6\,\leq\,R_{\rm GC}\,\leq\,11.5$\,kpc and Region\,C of $11.5\,\leq\,R_{\rm GC}\,\leq\,22$\,kpc.
Region\,A is not covered by our RC sample, thus we only determine the radial gradients of Regions\,B and C for the aforedefined individual height  bins.
The results are presented in Fig.\,9.
The Figure shows that, close to the midplane ($|Z|\,<\,1.0\,$kpc), the gradients of Region C are flatter than those of Region B. 
For the upper disk ($|Z|\,\geq\,1.0\,$kpc), the radial gradients found for  Regions B and C become  comparable.
Fig.\,10 plots the  radial gradients measured Regions B and C as a function of $|Z|$.
For Region B, the radial gradient flattens with increasing $|Z|$.
It has a negative gradient of $-0.09$\,kpc\,dex$^{-1}$ at midplane and a marginally  positive of gradient $0.02$\,kpc\,dex$^{-1}$ at $\sim$\,2.5\,kpc away from the midplane.
 For Region C,  the gradient is essentially constant, value of  $-0.014\,\pm\,0.008$\,dex\,kpc$^{-1}$.

\begin{figure}
\centering
\includegraphics[scale=0.40,angle=0]{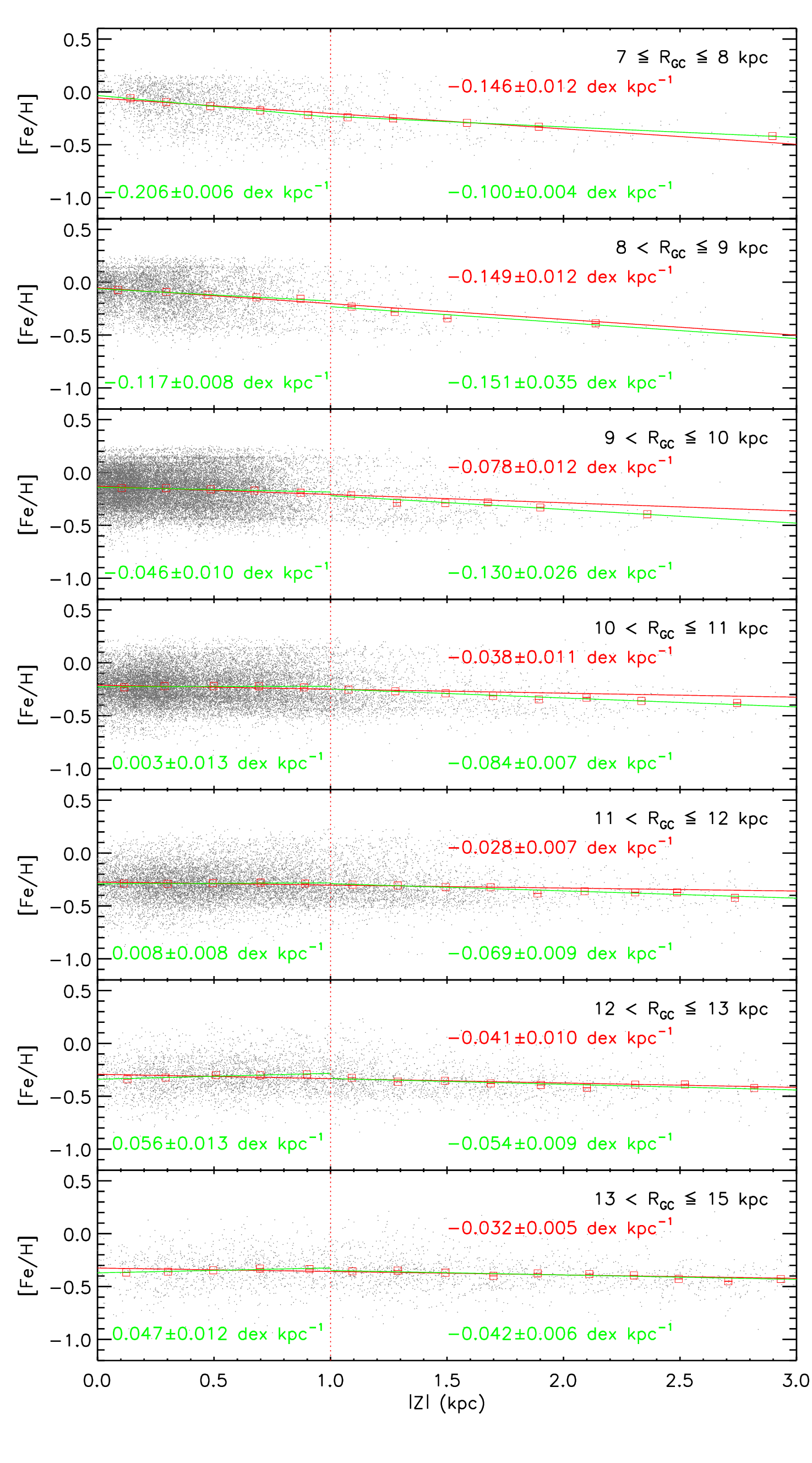}
\caption{Distributions of  metallicities, plotted  as a function of $|Z|$, deduced for the individual annuli of  $R_{\rm GC}$ from the RC sample. 
The red dashed lines divide the disk  into two regimes: the lower ($|Z| \leq 1$\,kpc) and upper  ($|Z| > 1$\,kpc) disk.
Red boxes gives the mean metallicities in the individual vertical bins.
The linear fits to the data  of the lower and upper disks  are shown by green lines, while those for the whole disk height are plotted in red.
The gradients yielded by the fits are also marked in  each plot, with the color code as for the fit.}
\end{figure}

\begin{figure}
\centering
\includegraphics[scale=0.5,angle=0]{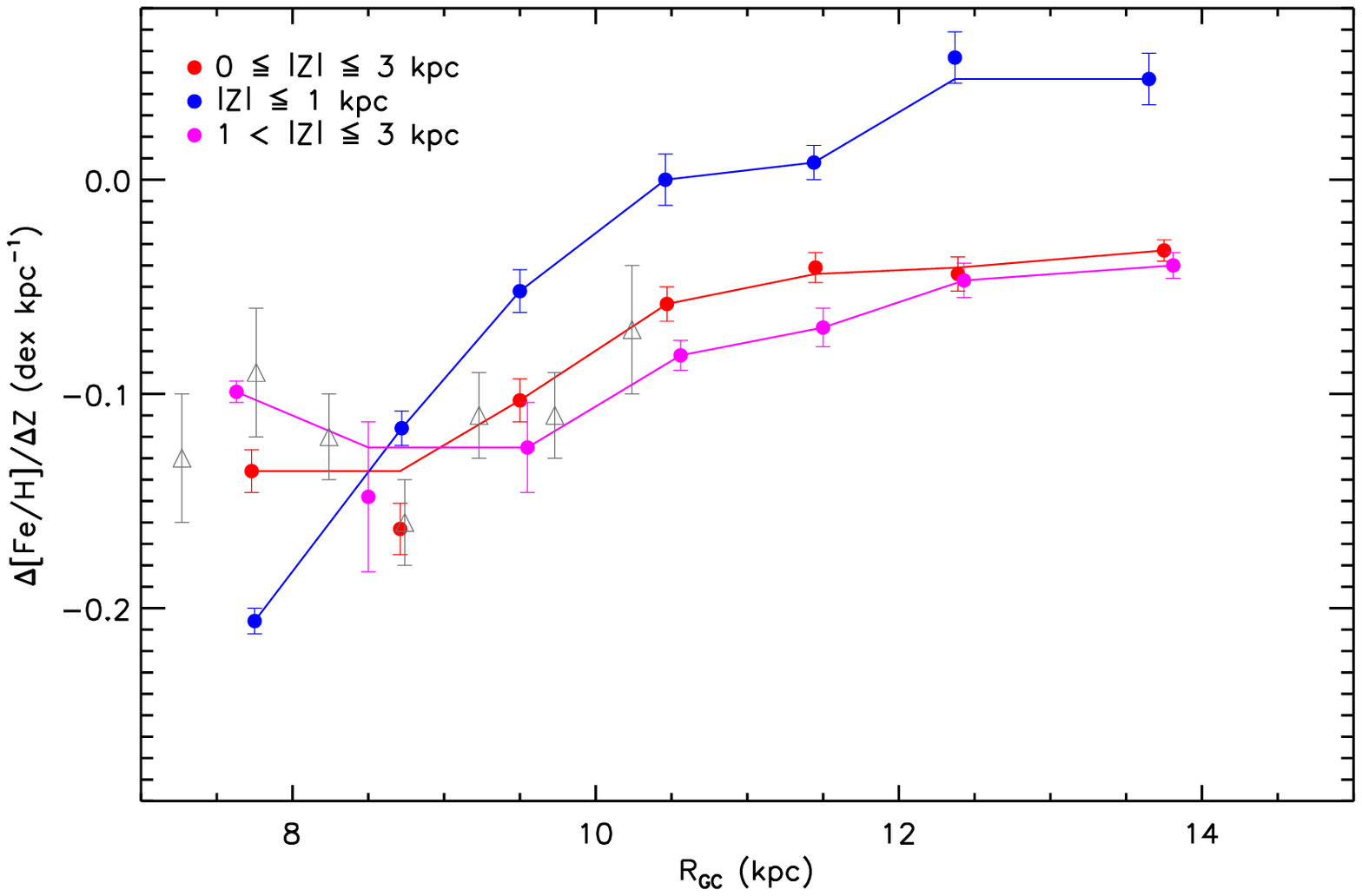}
\caption{Vertical gradients, $\Delta$[Fe/H]/$\Delta$Z, plotted as a function of $R_{\rm GC}$, deduced for different ranges of disk height $|Z|$ (Red:  $|Z|\,\leq\,3$\,kpc; Blue: $|Z|\,\leq\,1$\,kpc; and Mangeta:  $1\,<\,|Z|\,\leq\,3$\,kpc).
The lines connecting the data have been  smoothed over  3 adjacent points.
The grey triangles are measurements from Carrell et al. (2012) using SEGUE F/G/K dwarfs of $1\,\leq\,|Z|\,\leq\,3$\,kpc.
}
\end{figure}

\subsection{Vertical metallicity gradients}
To study the spatial variations of vertical gradients and also to minimize the effects of  radial and vertical gradients  on the determinations of vertical gradient, we divide the RC sample into annuli of width  1 kpc in the $R_{\rm GC}$ direction, except for the most distant annulus of 13\,$<$\,$R_{\rm GC}\,\leq\,15$\,kpc. 
To derive the gradients, we calculate the mean metallicities of the individual height bins.
The binsize in height $|Z|$ is allowed to vary but set to be no less than 0.2\,kpc  and each bin  contain s no less than  50 stars.
Again, a simple straight line is used to  fit  the data, and the gradient is given by  the slope of the fit. 

The results  are presented in Fig.\,11.
The derived vertical gradients as a function of $R_{\rm GC}$ are  presented  in Fig.\,12.
It shows that  the vertical gradient flattens with increasing $R_{\rm GC}$.
To better understand the spatial variations of vertical gradient, especially if there is any difference between the thin and thick disks, we further calculate the vertical gradients for the  lower ($|Z|\,\leq\,1$\,kpc) and upper disk ($1\,<\,|Z|\,\leq\,3$\,kpc) separately.
{ The separation is chosen to be 1\,kpc, where is the transition zone between the thin and thick disks given their scale heights (300\,pc for the thin disk and 900\,pc for the thick disk; Juri{\'c} et al. 2008).}  
As  Fig.\,15 shows, we find that the vertical gradient of the lower disk flattens quicker than that of the upper disk.
At $R_{\rm GC}\,>\,10.5$\,kpc, the vertical gradients of the lower disk diminish to zero and even become positive, while those of the upper disk remain negative and change only slowly with $R_{\rm GC}$.

\section{Discussions}
The LSS-GAC DR2 RC sample presented here allows us  to investigate the radial and vertical gradients over a large volume  of the Galactic disk, especially of its outer region.
In the following subsections, we will compare our results  to previous determinations  and discuss their implications for the  Galactic disk formation and evolution.
  
\subsection{Radial metallicity gradients}
%For  the inner regime of the disk, we only have  one measurement from the oldest age bin of the GCS sample. 
%The stars in this region are mostly kinematic thick disk stars.
%The positive gradient measured in this region is not unexpected, considering the  positive gradient  recently found for [Fe/H] as a function of   $V_{\phi}$\footnote{  $V_{\phi}$  is equivalent to  $R_{\rm g}$, based on the definition of $R_{\rm g}$.} for thick disk stars (e.g. Lee et al. 2011; Kordopatis et al. 2011).
%In addition, as  Fig.\,9 shows, the data points in  Region\,B are in  general described by a straight line of a single slope, the mean metallicities of the inner radial annuli ($\leq$\,8\,kpc) show some deviation from this slope and tend to to exhibit a flat slope even for height bins close to the midplane, which is also found recently by Hayden et al. (2014) using APOGEE red giants.
%This flattening gradient in the inner region is possibly due to the smaller scale length of the thick disc with respect to the thin disc.
%In this case,  more thick disk stars would be concentrated towards smaller projected Galactocentric radius, creating the flat gradient in the inner region assuming a flat radial gradient for the thick disk stars (see below).
%The possibility of the existence of a central bar that mix stars with different populations also can not be ruled out.  
The gradients near the solar circle (i.e. Region\,B) are well determined  in the current study given the   large number of  RC stars with accurate distances assembled here.
Near the midplane ($|Z|$\,$\leq\,0.1$\,kpc), the radial gradient measured from our RC sample has a value of  $0.082\,\pm\,0.003$\,dex\,kpc$^{-1}$, which is in excellent agreement with the result of $0.09\,\pm\,0.01$ \,dex\,kpc$^{-1}$ obtained by B14, also using  RC stars  near the midplane ($|Z|\,\leq\,0.05$\,kpc) with a similar coverage of $R_{\rm GC}$.  
Our results show that the radial gradient flattens with increasing disk height $|Z|$. 
A similar trend has been found by a number of recent studies (e.g. Cheng et al. 2012; Anders et al. 2014; Boeche et al. 2014; Hayden et al. 2014).
In Fig.\,10 we have also overplotted the measured radial gradients measured at different heights $|Z|$ by Anders et al. (2014) using APOGEE red giants and by Boeche et al. (2014) using RAVE RC stars.
Their measurements are generally in good agreement with our results.
{If one assumes that thin disk stars concentrate more closely to the midplane than thick disk stars, the trend of variations of  radial gradient with $|Z|$ may suggest that  the thin disk has a  negative  radial metallicity gradient while the thick disk has no radial metallicity gradient.}
{In addition, as  Fig.\,9 shows, the data points in  Region\,B are in  general well described by a straight line of a single slope, while the mean metallicities at the inner radial annuli ($\leq$\,8\,kpc) show some deviations from this slope and tend to to exhibit a flatter slope even for height bins close to the midplane.
Similar results have also been found recently by Hayden et al. (2014) using APOGEE red giants as tracers.
This flattening of gradient in the inner region is possibly caused by the existence of a central bar that mixes stars of different populations, as suggested by  Hayden et al. (2014).}

For the outer disk (i.e. Region\,C), as described earlier, the previous studies are not in conclusive with regard to whether the gradient is flatter  than near the solar circle owing to the poor  sampling  of the available samples.
In the current study, with a large number of RC stars sampling the outer disk, we are able to confirm that the metallicities in the outer Region\,C is essentially constant, and the gradients derived are indeed less steep than those found for the solar circle.
The  different trends of radial gradients as a function of $|Z|$  between Region\,B and C, as shown in Fig.\,10, may suggest that these two parts of the disk may have quite {different evolution paths}, resulted from, for example, infalling gas enriched in the halo  (e.g. Chiappini et al. 2001), the presence of non-axisymmetric structures (e.g. a central bar, long-lived spiral arms; e.g. Scarano \& L\'epine 2013), or a merge event in the outer disk (e.g. Yong 2006).

\subsection{Vertical metallicity gradients}
The vertical gradients are poorly determined  in previous studies. 
Most of those  studies find a steep gradient near the solar circle, for example, $-0.23$\,dex\,kpc$^{-1}$ by Barta{\v s}i{\= u}t{\.e} et al. (2003), $-0.30$\,dex\,kpc$^{-1}$ by Chen et al. (2003) and $-0.29$\,dex\,kpc$^{-1}$ by Marsakov\,\&\,Borkova (2006).
Those values are all slightly steeper than our result near the solar circle. 
The discrepancies are possibly caused by the fact that the sample tracers used by those earlier studies cover a wide range of $R_{\rm GC}$ and the vertical gradients are deduced without binning the data in $R_{\rm RG}$. Thus the values derived may be affected by the presence of  radial gradient, the so-called  the correlation effects discussed earlier.
{ The different age  distributions  of tracers used in different studies could also be responsible for the discrepancies.}
For the same $R_{\rm GC}$ or $|Z|$ range, our results are in excellent agreement with the previous studies.
As  Fig.\,12 shows, the vertical gradients derived at  different $R_{\rm GC}$ for  the upper disk from our RC sample  are comparable to the measurements by Carrel et al. (2012) using SEGUE F/G/K dwarfs of heights $1\,\leq\,|Z|\,\leq\,3$\,kpc.

In the current study, we find for the first time that  the vertical gradients of lower disk ($0 < |Z| < 1$\,kpc) flatten  with increasing $R_{\rm GC}$ quicker with $R_{\rm GC}$ than that of upper disk ($ 1 < |Z| < 3$\,kpc). { The result should provide important constraints on the formation scenarios of thin and thick disks.} 

%\subsection{Implications of the Galactic disk(s) formation and evolution}

%Radial gradient break: chemical evolution history

%temporal variations:

%radial migration

\section{Conclusions}
With improved  \logg\, measurements obtained with a  KPCA method trained by accurate asteroseismic data from the LAMOST\,--\,$Kepler$ fields, we have assembled hitherto the largest clean sample of RC stars, selected from the LSS-GAC DR2. 
The sample contains  over 70\,000 objects of intermediate ages, with a typical distance uncertainty  of $5$\,--\,$10$\,\%.
The sample stars cover a significant  disk volume of projected Galactocentric radii $7$\,$\leq$\,$R_{\rm GC}$\,$\leq$\,$14$\,kpc and heights from the Galactic midplane $0$\,$\leq$\,$|Z|$\,$\leq$\,$3$\,kpc.
With this RC sample, we have measured the radial and vertical metallicity gradients and studied their spatial variations.
%With other two sample from the literature with accurate age determinations, we study the temporal variations of the radial metallicity gradients.
The main conclusions of the current study are as follows:

1.\,Both the radial and vertical gradients are negative across much of the disk probed. The gradients also show significant spatial variations;

2.\,Near the solar circle ($7$\,$\leq$\,$R_{\rm GC}$\,$\leq$\,$11.5$\,kpc), the radial gradients flattens as one moves away from the Galactic plane;

3.\,In the outer disk ($11.5$\,$\leq$\,$R_{\rm GC}$\,$\leq$\,$14$\,kpc), the radial metallicity gradients do not show clear variations with  $|Z|$, and have an essentially constant value of  $-0.014\,$\,dex\,kpc\,$^{-1}$ much shallower than near the solar circle. This suggests that the outer disk may have a very {different evolution path};

%4.\,The gradient also shows significant temporal variations.
%The gradient show a monotone decreasing at the early epochs of disk formation and then flattening to a nearly constant value of $\sim$\,$-0.05$\,dex\,kpc\,$^{-1}$ at ages younger than  $\sim$\,4 Gyr.
%This result is compatible to the findings of Paper\,I and suggests the thick disk may formed at age older than $10$ Gyr with a flat radial gradient.

4.\,The vertical gradients also show significant spatial variations, flattening with increasing  $R_{\rm GC}$.
More interestingly, for the first time, we find the vertical gradients of the lower disk ($0\,<\,|Z|\,\leq\,1$\,kpc) flattens with  $R_{\rm GC}$ quicker than those of the upper disk ($1\,<\,|Z|\,\leq\,3$\,kpc).

For the moment, the chemical or chemo-dynamical  models can not fully predict the spatial variations of metallicity gradients found in the current work. 
Our results suggest that processes such as radial migration, non-axisymmetric perturbation, and galaxy mergers may have  played a significant role in shaping the Galactic disk(s).
The LSS-GAC survey is an ongoing project and will last until 2017.
With more observations  becoming available, the size and spatial coverage of our sample will continue to  expand.
In addition, additional information, especially the [$\alpha$/Fe] ratios, will become available with the new version of LSP3.
As the survey progresses, the LSS-GAC survey will in no doubt continue to provide further, stronger constraints on the chemical enrichment history of the Galactic disk(s),  and shed light on the formation and evolution of the Galaxy.

 \section*{Acknowledgements} 
 This work is supported by National Key Basic Research Program of China 2014CB845700 and  National Natural Science Foundation of China under grant number 11473001.
H.Y. thanks Professor Jian-Ning Fu for providing the LAMOST data of Kepler fields.
 
 The Guoshoujing Telescope (the Large Sky Area Multi-Object Fiber Spectroscopic Telescope, LAMOST) is a National Major Scientific Project built by the Chinese Academy of Sciences. Funding for the project has been provided by the National Development and Reform Commission. LAMOST is operated and managed by the National Astronomical Observatories, Chinese Academy of Sciences.

\appendix
\section{}
{The number of PCs adopted in our current KPCA model, i.e. $N_{\rm PC}\,=25$, is a tradeoff of the following two considerations.
Firstly,  a sufficient number of PCs is required to construct a tight relation between asteroseismic \logg\, and LAMOST blue-arm spectral features.
As Fig.\,A1 shows, when $N_{\rm PC}$ is small, the fit residuals show a large systematics.
When $N_{\rm PC}\,\ge\,25$, the fit residuals no longer show obvious trend as the atmospheric parameters vary and the standard deviations of the fit residuals decrease slowly with increasing $N_{\rm PC}$.
On the other hand, a value of $N_{\rm PC}$ that is too large will lead to biased estimates of $\log\,g$ under low spectral S/N's.
As Fig.\,A2 shows, the systematics  become quite significant  for spectral S/N's smaller than 20 if $N_{\rm PC} \ge 30$. 
Hence, we have adopted in the current work  $N_{\rm PC}\,=\,25$ for KPCA model in constructing the relation between asteroseismic \logg\, and LAMOST blue-arm spectral features.}

\begin{figure}
\centering
\includegraphics[scale=0.55,angle=0]{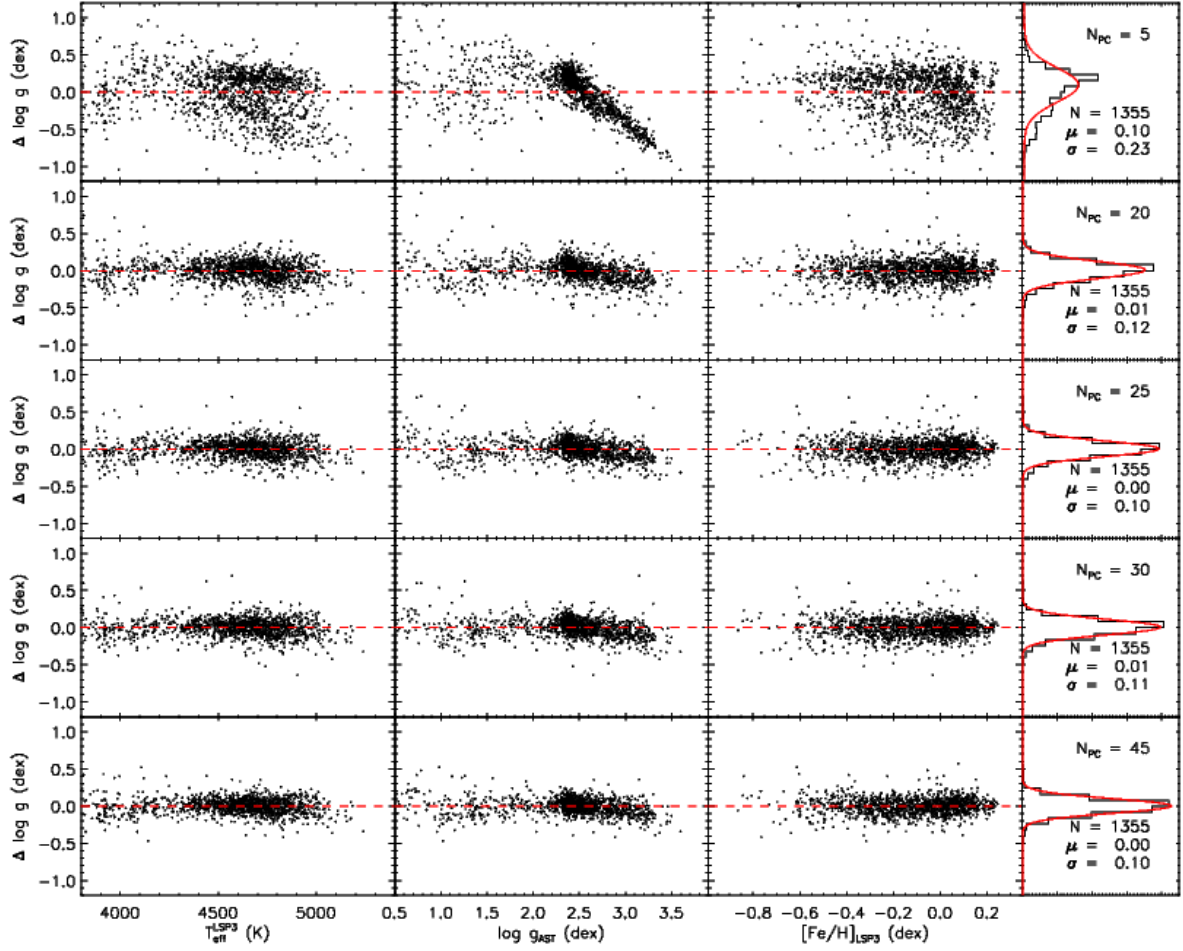}
\caption{Distributions of values of residual, log\,$g_{\rm KPCA}-$log\,$g_{\rm AST}$, of the training sample, as a function of LSP3 $T_{\rm eff}$ and [Fe/H] and of asteroseismic log\,$g$ for different $N_{\rm PC}$ assumed.
The last panel in each subplot shows a histogram distribution of the residuals (black line).
Also overplotted in red is a Gaussian fit to the distribution.
The mean  and dispersion of the fit, as well as the number of stars used, are marked. }
\end{figure}

\begin{figure}
\centering
\includegraphics[scale=0.5,angle=0]{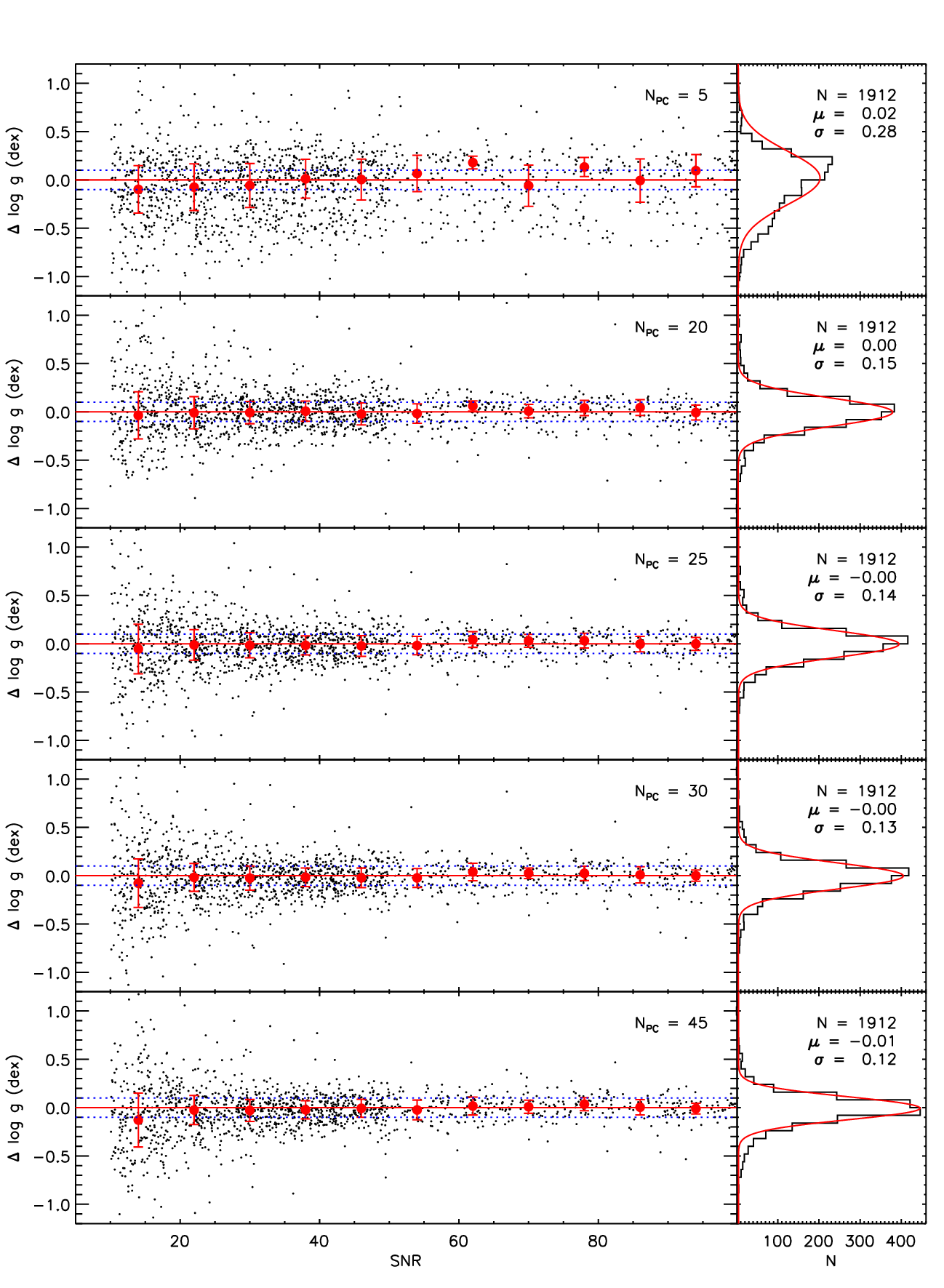}
\caption{Differences of  KPCA log g estimates and asteroseismic values for  the control sample, plotted as a function of spectral  S/N (4650\,\AA) for different $N_{\rm PC}$ assumed. 
The blue dashed lines indicate differences $\Delta$\,log\,$g$\,$=$\,$\pm$\,0.1\,dex.
The means and standard deviations of the differences in the individual S/N (4650\,\AA) bins (binsize\,$=$\,8 ) are oveplotted with dots and error bars, respectively.}
\end{figure}


\begin{thebibliography}{}
%-------A------------
\bibitem[Ahn et al.(2014)]{2014ApJS..211...17A} Ahn, C.~P., Alexandroff, R., Allende Prieto, C., et al.\ 2014, \apjs, 211, 17
\bibitem[Allende Prieto et al.(2006)]{2006ApJ...636..804A} Allende Prieto, C., Beers, T.~C., Wilhelm, R., et al.\ 2006, \apj, 636, 804
\bibitem[Allende Prieto et al.(2008)]{2008AN....329.1018A} Allende Prieto, C., Majewski, S.~R., Schiavon, R., et al.\ 2008, Astronomische Nachrichten, 
329, 1018 

\bibitem[Anders et al.(2014)]{2014A&A...564A.115A} Anders, F., Chiappini, C., Santiago, B.~X., et al.\ 2014, \aap, 564, 115
\bibitem[Andrievsky et al. (2002)]{2002A&A...381...32A} Andrievsky, S.~M., Kovtyukh, V.~V. \& Luck, R.~E. et al. 2002a, A\&A, 381, 32
\bibitem[Andrievsky et al. (2002b)]{2002A&A...384..140A} Andrievsky, S.~M., Bersier, D. \& Kovtyukh, V.~V. et al. 2002b, A\&A, 384, 140
\bibitem[Andrievsky et al. (2002c)]{2002A&A...392..491A} Andrievsky, S.~M., Kovtyukh, V.~V. \& Luck, R.~E. et al. 2002c, A\&A, 392, 491
\bibitem[Andrievsky et al. (2004)]{2004A&A...413..159A} Andrievsky, S.~M., Luck, R.~E., Martin, P. \& L\'epine, J.~R.~D., 2004, A\&A, 413, 159

%-------B------------
\bibitem[Balser et al. (2011)]{2011ApJ...738...27B}Balser, D.~S., Rood, R.~T., Bania, T.~M. \& Anderson, L.~D., 2011, ApJ, 738, 27
\bibitem[Barta{\v s}i{\= u}t{\.e} et al.(2003)]{2003BaltA..12..539B} Barta{\v s}i{\= u}t{\.e}, S., Aslan, Z., Boyle, R.~P., et al.\ 2003, BaltA, 12, 539 

\bibitem[Belkacem et al.(2011)]{2011A&A...530A.142B} Belkacem, K., Goupil, M.~J., Dupret, M.~A., et al.\ 2011, \aap, 530, 142
\bibitem[Bilir et al.(2012)]{2012MNRAS.421.3362B} Bilir, S., Karaali, S., Ak, S., et al.\ 2012, \mnras, 421, 3362 
\bibitem[Bienaym{\'e} et al.(2014)]{2014A&A...571A..92B} Bienaym{\'e}, O., Famaey, B., Siebert, A., et al.\ 2014, \aap, 571, 92
\bibitem[Boeche et al.(2013)]{2013A&A...559A..59B} Boeche, C., Siebert, A., Piffl, T., et al.\ 2013, \aap, 559, 59
\bibitem[Boeche et al.(2014)]{2014A&A...568A..71B} Boeche, C., Siebert, A., Piffl, T., et al.\ 2014, \aap, 568, 71 
\bibitem[Borucki et al.(2010)]{2010Sci...327..977B} Borucki, W.~J., Koch, D., Basri, G., et al.\ 2010, \sci, 327, 977 
\bibitem[Bovy et al.(2014)]{2014ApJ...790..127B} Bovy, J., Nidever, D.~L., Rix, H.-W., et al.\ 2014, \apj, 790, 127
\bibitem[Bressan et al.(2012)]{2012MNRAS.427..127B} Bressan, A., Marigo, P., Girardi, L., et al.\ 2012, \mnras, 427, 127
\bibitem[Brown et al.(1991)]{1991ApJ...368..599B} Brown, T.~M., Gilliland, R.~L., Noyes, R.~W., \& Ramsey, L.~W.\ 1991, \apj, 368, 59 



%-------C------------
\bibitem[Cannon(1970)]{1970MNRAS.150..111C} Cannon, R.~D.\ 1970, \mnras, 150, 111
\bibitem[Carraro et al. (2007)]{2007A&A...476..217C}Carraro, G., Geisler, D., Villanova, S. et al. 2007, A\&A, 476, 217
\bibitem[Carrell et al.(2012)]{2012AJ....144..185C} Carrell, K., Chen, Y., \& Zhao, G.\ 2012, \aj, 144, 185 
\bibitem[Carrera \& Pancino(2011)]{2011A&A...535A..30C} Carrera, R., \& Pancino, E.\ 2011, \aap, 535, 30 
\bibitem[Casagrande et al.(2011)]{2011A&A...530A.138C} Casagrande, L., Sch{\"o}nrich, R., Asplund, M., et al.\ 2011, \aap, 530, 138 
\bibitem[Chabrier(2001)]{2001ApJ...554.1274C} Chabrier, G.\ 2001, \apj, 554, 1274 
\bibitem[Chen et al. (2003)]{2003AJ....125.1397C} Chen, L., Hou, J.~L. \& Wang, J.~J., 2003, AJ, 125, 1397
\bibitem[Chen et al. (2011)]{2011AJ....142..184C}Chen, Y.~Q., Zhao, G., Carrell, K., Zhao, J.~K., 2011, AJ, 142, 184
\bibitem[Cheng et al. (2012)]{2012ApJ...746..149C} Cheng, J.~Y., Rockosi, C.~M., Morrison, H.~L. et al. 2012, ApJ, 746, 149
\bibitem[Chiappini et al.(1997)]{1997ApJ...477..765C} Chiappini, C., Matteucci, F., \& Gratton, R.\ 1997, \apj, 477, 765 
\bibitem[Chiappini et al.(2001)]{2001ApJ...554.1044C} Chiappini, C., Matteucci, F., \& Romano, D.\ 2001, \apj, 554, 1044 
\bibitem[Creevey et al.(2013)]{2013MNRAS.431.2419C} Creevey, O.~L., Th{\'e}venin, F., Basu, S., et al.\ 2013, \mnras, 431, 2419 
\bibitem[Costa et al. (2004)]{2004A&A...423..199C}Costa, R.~D.~D., Uchida, M.~M.~M., Maciel, W.~J., 2004, A\&A, 423, 199
\bibitem[Cui et al. (2012)]{Cui2012}Cui X.-Q. et al., 2012, RAA, 12, 1197

%-------D------------
\bibitem[Daflon \& Cunha (2004)]{2004ApJ...617.1115D}Daflon, S. \& Cunha, K., 2004, ApJ, 617, 1115
\bibitem[Daflon et al.(2009)]{2009AJ....138.1577D} Daflon, S., Cunha, K., de la Reza, R., Holtzman, J., \& Chiappini, C.\ 2009, \aj, 138, 1577
\bibitem[De Cat et al. (2014)]{2014arXiv1411.0913D} De Cat P., Fu J.-N., Yang, X.-H. et al. 2014, preprint, arXiv:1411.0913
\bibitem[Deng et al. (2012)]{2012RAA...12..735D} Deng, L.-C., Newberg, H.~J., Liu, C., et al.\ 2012, RAA, 12, 735
\bibitem[Deharveng et al. (2000)]{2000MNRAS.311..329D}Deharveng, L., Pe$\tilde{\rm n}$a, M., Caplan, J. \& Costero, R., 2000, MNRAS, 311, 329

%-------E------------

\bibitem[Epstein et al.(2014)]{2014ApJ...785L..28E} Epstein, C.~R., Elsworth, Y.~P., Johnson, J.~A., et al.\ 2014, \apjl, 785, LL28 
\bibitem[ESA(1997)]{1997ESASP1200.....E} ESA 1997, ESA Special Publication, 1200


%-------F------------
\bibitem[Re Fiorentin et al.(2007)]{2007A&A...467.1373R} Re Fiorentin, P., Bailer-Jones, C.~A.~L., Lee, Y.~S., et al.\ 2007, \aap, 467, 1373
\bibitem[Friel et al. (2002)]{2002AJ....124.2693F}Friel, E.~D., Janes, K.~A., Tavarez, M. et al. 2002, AJ, 124, 2693
\bibitem[Friel et al. (2010)]{2010AJ....139.1942F} Friel, E.~D., Janes, K.~A. \& Tavarez, M., 2010, AJ, 139, 1942
\bibitem[Frinchaboy et al.(2013)]{2013ApJ...777L...1F} Frinchaboy, P.~M., Thompson, B., Jackson, K.~M., et al.\ 2013, \apjl, 777, LL1 
\bibitem[Fu et al.(2009)]{2009ApJ...696..668F} Fu, J., Hou, J.~L., Yin, J., \& Chang, R.~X.\ 2009, \apj, 696, 668

%-------G------------
\bibitem[Genovali et al. (2013)]{2013A&A...554A.132G}Genovali, K., Lemasle, B., Bono, G. et al. 2013, A\&A, 554, 132
\bibitem[Genovali et al.(2014)]{2014A&A...566A..37G} Genovali, K., Lemasle, B., Bono, G., et al.\ 2014, \aap, 566, 37
\bibitem[Gummersbach et al. (1998)]{1998A&A...338..881G} Gummersbach, C.~A., Kaufer, A. \& Schaefer, D.~R. et al. 1998, A\&A, 338, 881

%-------H------------
\bibitem[Hayden et al. (2014)]{2014AJ....147..116H}Hayden, M.~R., Holtzman, J.~A., Bovy, J. et al. 2014, AJ, 147, 116
\bibitem[Henry et al. (2004)]{2004AJ....127.2284H}Henry, G.~W., Fekel, F.~C. \& Henry, S.~M., 2004, AJ, 127, 2284
\bibitem[Henry et al. (2010)]{2010ApJ...724..748H}Henry, R.~B.~C., Kwitter, K.~B., Jaskot, A.~E. et al. 2010, ApJ, 724, 748
\bibitem[Hou et al.(2000)]{2000A&A...362..921H} Hou, J.~L., Prantzos, N., \& Boissier, S.\ 2000, \aap, 362, 921 
\bibitem[Huber et al.(2014)]{2014ApJS..211....2H} Huber, D., Silva Aguirre, V., Matthews, J.~M., et al.\ 2014, \apjs, 211, 2 

%-------L------------
\bibitem[Laney et al.(2012)]{2012MNRAS.419.1637L} Laney, C.~D., Joner, M.~D., \& Pietrzy{\'n}ski, G.\ 2012, \mnras, 419, 1637 
\bibitem[Larson (1976)]{1976MNRAS.176...31L} Larson, R.~B. 1976, MNRAS, 176, 31
\bibitem[Lemasle et al. (2007)]{2007A&A...467..283L}Lemasle, B., Fran\c{c}ois, P., Bono, G. et al. 2007, A\&A, 467, 283
\bibitem[Lemasle et al. (2008)]{2008A&A...490..613L}Lemasle, B., Fran\c{c}ois, P., Piersimoni, A. et al. 2008, A\&A, 490, 613
\bibitem[Lemasle et al. (2013)]{2013A&A...558A..31L}Lemasle, B., Fran\c{c}ois, P., Genovali, K. et al. 2013, A\&A, 558, 31
\bibitem[Liu et al.(2014)]{2014arXiv1411.0235L} Liu, C., Fang, M., Wu, Y.,  et al.\ 2014, arXiv:1411.0235 
\bibitem[Liu et al. (2014)]{2014IAUS..310.321} Liu, X.-W., Yuan, H.-B., Huo, Z.-Y., et al.\ 2014, in Feltzing, S., Zhao, G., Walton, N., Whitelock, P., eds, Proc. IAU Symp. 298, Setting the scene for Gaia and LAMOST, Cambridge University Press, pp. 310-321, preprint (arXiv: 1306.5376)
\bibitem[Loebman et al. (2011)]{2011ApJ...737....8L} Loebman, S.~R., Ro$\check{s}$kar, R., Debattista, V.~P. et al. 2011, ApJ, 737, 8
\bibitem[Luck et al. (2003)]{2003A&A...401..939L} Luck, R.~E., Gieren, W.~P. \& Andrievsky, S.~M. et al. 2003, A\&A, 2003, 401, 939
\bibitem[Luck et al. (2006)]{2006AJ....132..902L}Luck, R.~E., Kovtyukh, V.~V., Andrievsky, S.~M. et al. 2006, AJ, 132, 902
\bibitem[Luck et al. (2011)]{2011AJ....142...51L}Luck, R.~E., Andrievsky, S.~M., Kovtyukh, V.~V. et al. 2011, AJ, 142, 51
\bibitem[Luck \& Lambert (2011)]{2011AJ....142..136L} Luck, R.~E. \& Lambert, D.~L., 2011, AJ, 142, 136
\bibitem[Luo et al.(2015)]{2015arXiv150501570L} Luo, A.-L.,  et al.\ 2015, arXiv:1505.01570 

%-------M------------
\bibitem[Maciel et al.(2005)]{2005A&A...433..127M} Maciel, W.~J., Lago, L.~G., \& Costa, R.~D.~D.\ 2005, \aap, 433, 127 
\bibitem[Magrini et al. (2009)]{2009A&A...494...95M}Magrini, L., Sestito, P., Randich, S. \& Galli, D., 2009, A\&A 494, 95
\bibitem[Magrini et al. (2010)]{2010A&A...523A..11M}Magrini, L., Randich, S., Zoccali, M. et al. 2010, A\&A, 523, 11
\bibitem[Martin et al.(2015)]{2015arXiv150305898M} Martin, R.~P., Andrievsky, S.~M., Kovtyukh, V.~V., et al.\ 2015, arXiv:1503.05898 
\bibitem[Marsakov \& Borkova(2006)]{2006AstL...32..376M} Marsakov, V.~A., \& Borkova, T.~V.\ 2006, AstL, 32, 376
\bibitem[Matteucci \& Francois(1989)]{1989MNRAS.239..885M} Matteucci, F., \& Francois, P.\ 1989, \mnras, 239, 885
\bibitem[Minchev et al.(2013)]{2013A&A...558A...9M} Minchev, I., Chiappini, C., \& Martig, M.\ 2013, \aap, 558, AA9
\bibitem[Moll{\'a} et al.(1997)]{1997ApJ...475..519M} Moll{\'a}, M., Ferrini, F., \& D{\'{\i}}az, A.~I.\ 1997, \apj, 475, 519
\bibitem[Moll{\'a} \& D{\'{\i}}az(2005)]{2005MNRAS.358..521M} Moll{\'a}, M., \& D{\'{\i}}az, A.~I.\ 2005, \mnras, 358, 521 

%-------J------------
\bibitem[Jacobson et al. (2008)]{2008AJ....135.2341J} Jacobson, H.~R., Friel, E.~D. \& Pilachowski, C.~A., 2008, AJ, 135, 2341
\bibitem[Jacobson et al. (2009)]{2009AJ....137.4753J} Jacobson, H.~R., Friel, E.~D. \& Pilachowski, C.~A., 2009, AJ, 137, 4753
\bibitem[Jacobson et al.(2011)]{2011AJ....141...58J} Jacobson, H.~R., Friel, E.~D., \& Pilachowski, C.~A.\ 2011a, \aj, 141, 58
\bibitem[Jacobson et al. (2011)]{2011AJ....142...59J} Jacobson, H.~R., Pilachowski, C.~A., \& Friel, E.~D. 2011b, AJ, 142, 59
\bibitem[Juri{\'c} et al.(2008)]{2008ApJ...673..864J} Juri{\'c}, M., Ivezi{\'c}, {\v Z}., Brooks, A., et al.\ 2008, \apj, 673, 864 

%-------K------------
\bibitem[Katz et al. (2011)]{2011A&A...525A..90K}Katz, D., Soubiran, C., Cayrel, R. et al. 2011, A\&A, 525, 90
\bibitem[Kordopatis et al. (2011)]{2011A&A...535A.107K}Kordopatis, G., Recio-Blanco, A., de Laverny, P. et al. 2011, A\&A, 535, 107
\bibitem[Kovtyukh et al. (2005)]{2005PASP..117.1173K}Kovtyukh, V.~V., Wallerstein, G. \& Andrievsky, S.~M. 2005, PASP
\bibitem[Kubryk et al. (2013)]{2013MNRAS.436.1479K}Kubryk, M., Prantzos, N. \& Athanassoula, E. 2013, MNRAS, 176, 31


%-------P------------
\bibitem[Paczy{\'n}ski \& Stanek(1998)]{1998ApJ...494L.219P} Paczy{\'n}ski, B., \& Stanek, K.~Z.\ 1998, \apjl, 494, L219
\bibitem[Pedicelli et al. (2009)]{2009A&A...504...81P}Pedicelli, S., Bono, G., Lemasle, B. et al. 2009, A\&A, 504, 81
\bibitem[Pedicelli et al. (2010)]{2010A&A...518A..11P}Pedicelli, S., Lemasle, B., Groenewegen, M. et al. 2010, A\&A, 518, 11
\bibitem[Perinotto \& Morbidelli (2006)]{2006MNRAS.372...45P}Perinotto, M. \& Morbidelli, L., 2006, MNRAS, 372, 45
\bibitem[Pinsonneault et al.(2014)]{2014ApJS..215...19P} Pinsonneault, M.~H., Elsworth, Y., Epstein, C., et al.\ 2014, \apjs, 215, 19
\bibitem[Portinari \& Chiosi(1999)]{1999A&A...350..827P} Portinari, L., \& Chiosi, C.\ 1999, \aap, 350, 827
\bibitem[Prantzos \& Boissier(2000)]{2000MNRAS.313..338P} Prantzos, N., \& Boissier, S.\ 2000, \mnras, 313, 338
\bibitem[Puzeras et al.(2010)]{2010MNRAS.408.1225P} Puzeras, E., Tautvai{\v s}ien{\.e}, G., Cohen, J.~G., et al.\ 2010, \mnras, 408, 1225 

%-------Q------------
\bibitem[Quireza et al. (2006)]{2006ApJ...653.1226Q}Quireza, C., Rood, R.~T., Bania, T.~M. et al. 2006, ApJ, 653, 1226

%-------R------------
\bibitem[Rees et al.(2000)]{2000A&A...355..759R} Rees, D.~E., L{\'o}pez Ariste, A., Thatcher, J., \& Semel, M.\ 2000, \aap, 355, 759 
\bibitem[Ren et al. (2015)]{Ren et al. 2015}Ren at al. 2015, in preparation

\bibitem[Roeser et al.(2010)]{2010AJ....139.2440R} Roeser, S., Demleitner, M., \& Schilbach, E.\ 2010, \aj, 139, 2440 
\bibitem[Rolleston et al. (2000)]{2000A&A...363..537R}Rolleston, W.~R.~J., Smartt, S.~J., Dufton, P.~L. \& Ryans, R.~S.~I., 2000, A\&A, 363, 537
\bibitem[Rudolph et al. (2006)]{2006ApJS..162..346R}Rudolph, A.~L., Fich, M., Bell, G.~R. et al. 2006, ApJS, 162, 346

%-------S------------
\bibitem[Scarano \& L{\'e}pine(2013)]{2013MNRAS.428..625S} Scarano, S., \& L{\'e}pine, J.~R.~D.\ 2013, \mnras, 428, 625 
\bibitem[Schlesinger et al. (2012)]{2012ApJ...761..160S} Schlesinger, K.~J., Johnson, J.~A., Rockosi, C.~M. et al. 2012, ApJ, 761, 160
\bibitem[Schlesinger et al. (2014)]{2014ApJ...791..112S}Schlesinger, K.~J., Johnson, J.~A., Rockosi, C.~M. et al. 2014, ApJ, 791, 112
\bibitem[Sch{\"o}nrich \& Binney(2009)]{2009MNRAS.396..203S} Sch{\"o}nrich, R., \& Binney, J.\ 2009, \mnras, 396, 203 
\bibitem[Scholkopf et al. 1998]{B. Scholkopf et al. 1998}Sch\"{o}lkopf B.,  Smola A., \& Muller K.R.,  1998, Neural Computation, 10, 1299
\bibitem[Siebert et al.(2011)]{2011MNRAS.412.2026S} Siebert, A., Famaey, B., Minchev, I., et al.\ 2011, \mnras, 412, 2026 
\bibitem[Singh et al.(1998)]{1998MNRAS.295..312S} Singh, H.~P., Gulati, R.~K., \& Gupta, R.\ 1998, \mnras, 295, 312
\bibitem[Skrutskie et al.(2006)]{2006AJ....131.1163S} Skrutskie, M.~F., Cutri, R.~M., Stiening, R., et al.\ 2006, \aj, 131, 1163
\bibitem[Smartt \& Rolleston(1997)]{1997ApJ...481L..47S} Smartt, S.~J., \& Rolleston, W.~R.~J.\ 1997, \apjl, 481, L47
\bibitem[Stanek \& Garnavich(1998)]{1998ApJ...503L.131S} Stanek, K.~Z., \& Garnavich, P.~M.\ 1998, \apjl, 503, L131
\bibitem[Stanghellini et al. (2006)]{2006ApJ...651..898S}Stanghellini, L., Guerrero, M.~A., Cunha, K. et al. 2006, ApJ, 651, 898
\bibitem[Stanghellini \& Haywood (2010)]{2010ApJ...714.1096S} Stanghellini, L. \& Haywood, M. 2010, ApJ, 714, 1096 
\bibitem[Steinmetz et al.(2006)]{2006AJ....132.1645S} Steinmetz, M., Zwitter, T., Siebert, A., et al.\ 2006, \aj, 132, 1645
\bibitem[Stello et al.(2013)]{2013ApJ...765L..41S} Stello, D., Huber, D., Bedding, T.~R., et al.\ 2013, \apjl, 765, LL41

%-------T------------
\bibitem[Tosi(1988)]{1988A&A...197...33T} Tosi, M.\ 1988, \aap, 197, 33 

%-------W------------
\bibitem[Williams et al.(2013)]{2013MNRAS.436..101W} Williams, M.~E.~K., Steinmetz, M., Binney, J., et al.\ 2013, \mnras, 436, 101

\bibitem[Wu et al.(2011)]{2011RAA....11..924W} Wu, Y., Luo, A.-L., Li, H.-N., et al.\ 2011, RAA, 11, 924

%-------X------------
\bibitem[Xiang et al. (2015a)]{Xiang et al. 2015} Xiang et al. 2015a, in preparation
\bibitem[Xiang et al. (2015b)]{2015MNRAS.448...90X} Xiang, M.-S., Liu, X.-W., Yuan, H.-B. et al.\ 2015a, MNRAS, 448, 90
\bibitem[Xiang et al. (2015c)]{2015MNRAS.448..822X} Xiang, M.-S., Liu, X.-W., Yuan, H.-B. et al.\ 2015b, MNRAS, 448, 822

%-------Y------------
\bibitem[Yanny et al. (2009)]{2009AJ....137.4377Y}Yanny, B., Rockosi, C., Newberg, H.~J. et al. 2009, AJ, 137, 4377
\bibitem[Yong et al. (2005)]{2005AJ....130..597Y}Yong, ., Carney, B.~W. \& Teixera de Almeida, M.~L., 2005, AJ, 130, 597
\bibitem[Yong et al. (2006)]{2006AJ....131.2256Y} Yong, D., Carney, B.~W., Teixera de Almeida, M.~L. \& Pohl, B.~L., 2006, AJ, 131, 2256
\bibitem[Yong et al. (2012)]{2012AJ....144...95Y}Yong, D., Carney, B.~W. \& Friel, E.~D., 2012, AJ, 144, 95
\bibitem[York et al.(2000)]{2000AJ....120.1579Y} York, D.~G., Adelman, J., Anderson, J.~E., Jr., et al.\ 2000, \aj, 120, 1579
\bibitem[Yuan et al. (2015)]{2015MNRAS.448..855Y}Yuan, H.-B., Liu, X.-W., Huo, Z.-Y. et al.\ 2015, MNRAS, 448, 855

%-------V------------
\bibitem[Vilchez \& Esteban(1996)]{1996MNRAS.280..720V} Vilchez, J.~M., \& Esteban, C.\ 1996, \mnras, 280, 720 

\bibitem[Zacharias et al.(2013)]{2013AJ....145...44Z} Zacharias, N., Finch, C.~T., Girard, T.~M., et al.\ 2013, \aj, 145, 44





\end{thebibliography}
\end{document}